\documentclass[preprintnumbers,floats,floatfix,amssymb,prd,onecolumn,superscriptaddress,nofootinbib]{revtex4}
\usepackage{amssymb,amsmath,tabu}
\usepackage{epsfig,hyperref}
\usepackage[svgnames]{xcolor}
\usepackage{pgf,tikz}
\usepackage{epstopdf}
\usepackage[british]{babel}
\usepackage{enumerate}
\usepackage{enumitem}
\usepackage{booktabs}
\usepackage{multirow}
\usepackage{makecell}
\usepackage{color}

\makeatletter
\DeclareRobustCommand*{\bfseries}{%
  \not@math@alphabet\bfseries\mathbf
  \fontseries\bfdefault\selectfont
  \boldmath
}
\makeatother

\def\be{\begin{equation}}
\def\ee{\end{equation}}
\def\beq{\begin{eqnarray}}
\def\eeq{\end{eqnarray}}

\makeatletter

\newcommand{\arXiv}[2][]{\href{http://arxiv.org/abs/#2}{\texttt{arXiv:#2\@ifempty{#1}{}{ [#1]}}}}
\makeatother

\begin{document}
\title{Accretion disks around naked singularities}

\author{Jun-Qi Guo}%
\email{sps{\_}guojq@ujn.edu.cn}
\affiliation{School of Physics and Technology, University of Jinan, Jinan 250022, Shandong, China}

\author{Pankaj S. Joshi}%
\email{psjcosmos@gmail.com}
\affiliation{International Center for Cosmology, Charusat University, Anand 388421, Gujarat, India}

\author{Ramesh Narayan}%
\email{rnarayan@cfa.harvard.edu}
\affiliation{Harvard-Smithsonian Center for Astrophysics, 60 Garden Street, Cambridge, MA 02138, USA}

\author{Lin Zhang}%
\email{linzhang2013@pku.edu.cn}
\affiliation{College of Mathematics and Statistics, Chongqing University, Chongqing 401331, China}

\date{\today}

\begin{abstract}
We investigate here the thermal properties of accretion disks in  a spacetime for some galactic density profiles in spherical symmetry. The matter distributions have a finite outer radius with a naked central singularity. The luminosities of the accretion disks for some density profile models are found to be higher than those for a Schwarzschild black hole of the same mass. The slopes for the luminosity distributions with respect to frequencies are significantly different, especially at higher frequencies, from that in the Schwarzschild black hole case. Such features may be used to distinguish black holes from naked singularities. The efficiencies for the conversion of the mass energy of the accreting gas into radiation and the strength of naked singularities are analyzed. The novel feature that we find is, the strength of the singularity is different depending on the profiles considered, and the stronger the singularity is, the higher is the efficiency for the accretion disk.
\end{abstract}
\maketitle

\section{Introduction}

The phenomena of gravitational collapse and related spacetime singularities are basic to several key considerations in relativistic astrophysics, apart from being crucial to gravitation theory and mathematical relativity\cite{Oppenheimer_1939,Penrose_1965,Hawking_1973,Joshi_1993a,Wald:1997wa,Krasinski_1997,Krolak:1999wk,Berger_2002,Joshi_2007,
Henneaux_2008,Clarke_2012,Belinski_2018}. At the late stage of their nuclear burning cycles, massive stars would undergo continual collapse under self-gravity, to form ultra-compact entities such as spacetime singularities and black holes. The recent gravitational waves detection by the LIGO-Virgo Collaborations~\cite{Abbott_2016,LIGO_Virgo}, and the image of the black hole in Galaxy Messier 87, captured by the Event Horizon Telescope Collaboration~\cite{Akiyama:2019cqa,EHT}, ensure that we can now meaningfully explore the very strong gravity regions in the Universe. In addition to black holes, naked singularities are another set of possible final states of gravitational collapse~\cite{Joshi_1993a,Eardley_1979,Christodoulou_1984,Ori_1987,Choptuik:1992jv,Christodoulou_1994,Brady:1994aq,Zhang:2015rsa}.

In this paper, we  investigate the observational signatures of the accretion disks in spacetime models for some galactic mass distributions, which have a finite outer radius and have a naked singularity at the center. Such an investigation may help us to understand the nature of ultra-compact objects at the galactic centers in terms of black holes and naked singularities. The spacetime near a black hole is quite different from that near a naked singularity. Such a difference allows us to differentiate a naked singularity from a black hole. In Ref.~\cite{RN_1304}, a Joshi-Malafarina-Narayan (JMN2) model was proposed and studied. For this model, the matter distribution has a finite outer boundary, the density inside is an inverse square function of the radius, and therefore it is singular at the center. (For the JMN1 model, see Ref.~\cite{RN_1106}.) The observational signatures of the accretion disk around a naked singularity in the JMN2 model were then studied. The disk for this model is much more luminous than the one in the black hole case. More interesting is that the spectra have noticeably different shapes at high frequencies. This feature is observationally testable and can be used to distinguish naked singularities from a Schwarzschild black hole. In Ref.~\cite{Shahidi:2020bla}, the differences for the properties of the accretion disks around Kerr black holes and Brans-Dicke-Kerr naked singularities were investigated.

While the inverse square density distribution in the above work was studied as a toy model, it is meaningful to extend the study to more general and physically more realistic density distributions in spherical symmetry. The observed luminosity distributions of many elliptical galaxies and bulges are well described by the empirical de Vaucouleurs profile~\cite{Vaucouleurs}, and some other profiles are also used. Some density profile models have been constructed to reproduce the de Vaucouleurs profile. Typical examples include the Jaffe model~\cite{Jaf}, the Hernquist model~\cite{Hernquist}, the Dehnen model~\cite{Dehnen} and the Navarro-Frenk-White (NFW) model~\cite{NFW}. In this paper, we compute the spectral luminosity distributions of the accretion disks around naked singularities of such models. It is found that, for some models, the mass energy of the accreting gas cannot be totally converted into radiation. The spectra have different shapes at high frequencies, which can be applied to observationally differentiate the models.

An interesting and important finding that we report here is, depending on the nature of the density distribution chosen, the strength of the naked central singularity varies and changes accordingly. Moreover, we find that this has an important bearing on the efficiency of the corresponding accretion disks. The strength of spacetime singularities is an important feature and a characteristic to classify and understand the physical nature of singularities~\cite{Tipler_1980,Joshi_2007,Ellis_1977,Tipler_1977,Krolak:1999wk}.  Ellis and Schmidt classified curvature singularities into weak and strong ones. A singularity is called gravitationally weak when some object remains intact as the singularity is approached, and strong otherwise~\cite{Ellis_1977}. Tipler defined a singularity to be strong if an object approaching the singularity is crushed to a zero volume~\cite{Tipler_1977}. In mathematical language, a singularity is called gravitationally strong if the volume (area) element defined by three (two) independent vorticity-free Jacobi vector fields along a timelike (null) geodesic from the singularity, goes to a vanishing value as the singularity is approached. Krolak proposed a weaker version of the strong singularity condition, which states that a geodesic terminates at a strong curvature singularity if the expansion, a solution of the Raychaudhuri equation, along the geodesic is negative. In this definition, $\lq$the strong curvature singularity is identified by the property that it focuses all the congruences of the null geodesics approaching it'~\cite{Krolak_1986}. Clarke and Krolak obtained the necessary and sufficient conditions for occurrence of strong curvature singularities as defined by Tipler and Krolak~\cite{Clarke_1985}. Nolan modified the definition of the strength of singularities slightly, such that a singularity is also called strong if the volume element has infinite limit~\cite{Nolan:1999tw}. Considering that Tipler's definition is based on volume element constructed by three Jacobi fields, rather than the behavior of the individual Jacobi fields, Ori extended the class of strong singularities so as to include the circumstances where any of the Jacobi fields is unbounded~\cite{Ori_2000}.

With the above definitions, the strength of some spacetime singularities have been investigated. Some of these cases are the following. (i) Regarding the Lemaitre-Tolman-Bondi (LTB) model describing spherical collapse of inhomogeneous dust, the shell-crossing singularity in this model was shown to be weak by Newman~\cite{Newman_1986} and Nolan~\cite{Nolan:1999tw}. Also, the central shell-focussing naked singularity investigated by Eardley and Smarr~\cite{Eardley_1979} and Christodoulou~\cite{Christodoulou_1984} was shown to be weak by Newman~\cite{Newman_1986}. Strong curvature singularities exist in self-similar LTB  spacetimes~\cite{Waugh_1988}. Rajagopal and Lake found that the shell-focusing singularities in the ingoing Vaidya spacetimes are strong~\cite{Rajagopal_1987}. The naked singularities formed in self-similar collapse of perfect fluids were shown to be strong~\cite{Lake_1988,Waugh_1989,Ori_1990,Joshi_1992}. The strength of singularities in general class of models for the LTB collapse was analyzed by Joshi and Dwivedi~\cite{Joshi_1993}. (ii) The singularities in the Schwarzschild, Friedmann-Robertson-Walker and Kasner solutions and in black hole formation by scalar collapse were shown to be strong by Burko~\cite{Burko:1997xa}. The central singularity in the continuous self-similar scalar collapse (Roberts' solution)~\cite{Roberts} was shown to be weak by Nolan~\cite{Nolan:1999tw}. (iii) Consider some test matter falling onto the Cauchy horizon of a Reissner-Nordstr\"{o}m or Kerr black hole. Taking into account the curvature's backreaction on the matter, Poisson and Israel found that the Cauchy horizon will be transformed into a mass-inflation curvature singularity~\cite{Poisson_1989,Poisson_1990}. Ori showed that this singularity is weak~\cite{Ori_1991,Ori_1992}. In Ref.~\cite{Chesler:2019tco}, it was argued that, based on the Price's Law, tails and the exponential blueshift, a spacelike singularity will be formed inevitably in spherical collapse of a charged scalar field. For a review on the strength of singularities, see Ref.~\cite{Krolak:1999wk}.
In this paper, we investigate the strength of the naked singularities in the density profile models. It is found that the singularities in these models have different strength properties, and there is a close connection between the strength of the singularities and the efficiencies for the accretion disks. The stronger the singularity is, the higher is the efficiency.

The paper is organized as follows. In Sec.~\ref{sec:framework}, we present the framework of the study. We investigate the thermal properties of the accretion disks for some density profile models in Sec.~\ref{sec:luminosity}. The strength of the naked singularities in the density profile models is studied in Sec.~\ref{sec:strength}. The results are summarized in Sec.~\ref{sec:summary}. In this paper, we set $G=c=h=k=\sigma=1$, where $G$ is the Newtonian gravitational constant, $c$ the speed of light, $h$ the Planck constant, $k$ the Boltzmann constant, and $\sigma$ the Stefan-Boltzmann constant.

\section{Framework\label{sec:framework}}
In this section, we describe the framework of computing the spectral luminosity distributions of the accretion disks, including the metric for a perfect fluid sphere, the radiative properties of the accretion disks, density profile models and numerics etc.

\subsection{Metric}
A static spherically symmetric equilibrium configuration with a naked singularity at the center can be obtained from a slowly evolving gravitationally collapsing perfect fluid cloud. We mimic a galaxy by such an object. The interior static metric of the sphere can have the following form~\cite{RN_1304}
\be ds^2=-e^{2\phi}dt^2+\frac{1}{D}dr^2+r^2 d\Omega^2=-A{dt^2}+B{dr^2}+r^2 d\Omega^2, \hphantom{dddd} r\le r_{b}. \label{metric}\ee
where $\phi$, $D$, $A(\equiv e^{2\phi})$ and $B(\equiv1/D)$ are functions of the coordinate $r$, and $r=r_{b}$ the outer boundary of the sphere. The energy-momentum tensor for the perfect fluid source is $T_{\mu}^{\nu}=\mbox{diag}(\epsilon,p,p,p)/(8\pi)$. Set the function $F$ as the double Misner-Sharp mass,
\be D=1-\frac{F}{r}.\label{define_F}\ee
Then the Einstein equations and the equation of motion for the perfect fluid generate
\begin{align}
F_{,r}&=\epsilon r^2,\\
\nonumber\\
\phi_{,r}&=\frac{pr^{3}+F}{2r(r-F)},\label{phi_r}\\
\nonumber\\
p_{,r}&=-(\epsilon+p)\frac{pr^{3}+F}{2r(r-F)},\label{p_r}
\end{align}
where the $(_{,r})$ denotes a derivative with respect to the coordinate $r$. The interior metric is matched to a Schwarzschild exterior on the outer boundary $r=r_{b}$,
\be D(r_b)=1-\frac{2M_{T}}{r_b}=1-\frac{F(r_b)}{r_b},\label{boundary_condition}\ee
where $M_{T}$ is the total gravitational mass of the sphere.

\subsection{Thermal properties of the accretion disks}
Assume that a test particle orbiting inside the matter \lq cloud\rq~of the sphere does not interact with the material of the cloud, but is only affected by the gravity from the cloud. For simplicity, we set the total gravitational mass of the cloud to $1$. Moreover, considering that the metric is spherically symmetric, we choose the coordinate $\theta=\pi/2$ such that the geodesic trajectory of the test particle stays in the equatorial plane. In the circular geodesics circumstance, the energy per unit mass $E$, angular momentum per unit mass $L$ and angular velocity $\omega$ for the test particle are~\cite{RN_1304}
\be E^2=\frac{2A^2}{2A-A_{,r}r},\label{E_accretion}\ee
\be \frac{L^2}{r^2}=\frac{A_{,r}r}{2A-A_{,r}r},\label{L_accretion}\ee
\be \omega^2=-\frac{g_{tt,r}}{g_{\varphi\varphi,r}}=\frac{A_{,r}}{2r}.\label{omega_angular_frequency}\ee

We only consider the case in which the outer boundary of the sphere $r_b$ is outside the innermost stable circulate orbit (ISCO), $r_{\scriptsize\mbox{ISCO}}=6M_{T}$, such that the accretion disk extends with no break form large radius to the singularity~\cite{RN_1304}. The radiative properties of the disks can be calculated according to the relations obtained in Refs.~\cite{Novikov,Page}. As discussed in Ref.~\cite{RN_1304}, it is promising to use the spectral luminosity distribution to distinguish the density profile models from the Schwarzschild black hole model.

Assume that each local patch of an accretion disk radiates as a blackbody. We define a characteristic temperature $T_*$ by $\sigma T_*^4 \equiv \dot{m}c^2/[4\pi(GM_T/c^2)^2]$, where $\dot{m}$ is the rest mass accretion rate and is assumed to be constant, and $\sigma$ the Stefan-Boltzmann constant. Denote $\mathcal{F}(r)$ as the the radiative flux (energy per unit area per unit time) in the local frame of the accreting fluid. See Eq.~(\ref{flux}). Then the local blackbody temperature of the radiation is $T_{\rm BB}(r)=[{\mathcal F}(r)]^{1/4}T_*$. This radiation is transformed by gravitational and Doppler redshifts, and the transformation is dependent on the orientation of the observer with respect to the disk axis. For simplicity, we only consider the case where the observer stays on the disk axis. So the corresponding characteristic redshift $z$ is $1+z(r)=[-(g_{tt}+\omega^2 g_{\varphi\varphi})]^{-1/2}$. We also assume that the radiation emitted at radius $r$ has a temperature at infinity, independent of direction, $T_\infty(r)=T_{\rm BB}(r)/(1+z)$. Then the spectral luminosity distribution ${\cal L}_{\nu,\infty}$ observed by a face-on observer at infinity can be approximately written as
\be
\nu {\cal L}_{\nu,\infty} = \frac{15}{\pi^4} \int_{r_{\rm inner}}^\infty \left(\frac{d{\cal L}_\infty}{d\ln r}\right)\frac{(1+z)^4 (h\nu/kT_*)^4/\mathcal{F}}
{\exp[(1+z)(h\nu/kT_*)/{\mathcal F}^{1/4}]-1}\, d\ln r,
\label{integral_L}
\ee
where
\be
\frac{d{\cal L}_\infty}{d\ln r} = 4\pi r \sqrt{-g} E {\mathcal F},
\label{dLdr}
\ee
\be
\mathcal{F}(r)=-\frac{\dot{m}}{4\pi \sqrt{-g}}\frac{\omega_{,r}}
{(E-\omega L)^2}\int^r_{r_{\rm inner}}(E-\omega L)L_{,{\tilde{r}}}d{\tilde{r}}.
\label{flux}
\ee
$g$ is the determinant of the metric of the three-sub-space $(t,r,\varphi)$, $g(r)=-Ar^{2}/D$. $r_{\rm inner}$ is the radius of the inner edge of the accretion disk and also of the inner most stable circular orbit. $r_{\rm inner}$ is set to zero for the density profile models and to $6M_{\scriptsize T}$ for the Schwarzschild black hole model.

The efficiency for the conversion of the mass energy of the accreting gas into radiation can be obtained by integrating ${\cal L}_{\nu,\infty}$ over the frequency,
\be \mbox{Efficiency}=\int_{0}^{\infty}{\cal L}_{\nu,\infty}d\nu.\label{efficiency}\ee

\subsection{Density profile models of the galaxies\label{sec:models}}
\begin{itemize}[fullwidth,itemindent=0em]
  \item JMN2 model~\cite{RN_1304}. For the JMN2 model, inside the sphere, the density profile $\epsilon$, mass function $F$, metric component $g_{tt}=e^{2\phi}$ and pressure $p$ can be described as below:
  \be \epsilon=\frac{M_0}{r^2},\label{density_JMN2}\ee
  \be F=M_{0}r, \label{F_JMN2}\ee
  \be e^{2\phi}=(\alpha r^{1-\lambda}-\beta r^{1+\lambda})^2,\label{A_JMN2}\ee
  \be p=\frac{1}{2-\lambda^2}\frac{1}{r^2}\left[\frac{(1-\lambda)^2\alpha-(1+\lambda)^2\beta r^{2\lambda}}{\alpha-\beta r^{2\lambda}}\right],\label{p_JMN2}\ee
  where
  \be M_0=\frac{1-\lambda^2}{2-\lambda^2},\hphantom{dddd} \alpha=\frac{(1+\lambda)^2r_b^{\lambda-1}}{4\lambda\sqrt{2-\lambda^2}},
  \hphantom{dddd} \beta=\frac{(1-\lambda)^2r_b^{-\lambda-1}}{4\lambda\sqrt{2-\lambda^2}}, \hphantom{dddd} r_b=\frac{2M_{T}}{M_0}.\label{JMN2_parameters}\ee
  In this paper, we use $\lambda=1/\sqrt{2}$ and $\lambda=2/\sqrt{5}$.

  \item Dehnen model~\cite{Dehnen}. For the Dehnen model, there are
  \be \epsilon=\frac{(3-\gamma)M_{0}r_{0}^2}{r^{\gamma}(r+r_{0})^{4-\gamma}},\label{density_Dehnen}\ee
  \be F(r)=M_{0}r_{0}\left(\frac{r}{r+r_{0}}\right)^{3-\gamma}.\label{F_Dehnen}\ee
  The outer boundary of the sphere is
  \be r_b=r_0\left\{\left[1-\left(\frac{2M_{T}}{M_{0}r_{0}}\right)^{1/(3-\gamma)}\right]^{-1}-1\right\}.\label{boundary_Dehnen}\ee
Note that in the original definition of $\epsilon$ by Dehnen~\cite{Dehnen}, the power of $r_0$ in the numerator of the right hand side of Eq.~(\ref{density_Dehnen}) is $1$. In this paper, we set it to $2$ such that the expressions of $\epsilon$ and $F(r)$ are consistent with those in the JMN2 model. To avoid apparent horizon formation, we require $F(r)<r$ at all radii. Applying this requirement to small radii $r\ll1$, one obtains $\gamma\le2$. In this paper, we consider the range of $1\le\gamma\le2$. $\gamma=1$ and $\gamma=2$ correspond to the Hernquist and Jaffe models, respectively. Equation~(\ref{boundary_Dehnen}) implies that in order to make sure that the radius of the sphere $r_b$ is positive, there should be $M_0>2M_{T}/r_0$. Throughout the paper, we set $r_0=10$.

  \item NFW model~\cite{NFW}. For the NFW model, there are
  \be \epsilon=\frac{M_{0}r_{0}}{r(r+r_{0})^{2}},\ee
  \be F(r)=M_{0}r_{0}\left[\ln\left(\frac{r+r_0}{r_0}\right)-\frac{r}{r+r_0}\right].\ee
  The NFW model was originally proposed for the mass distribution of the cold dark matter halos rather than for galaxies. However, considering that this model is close to the Dehnen model near the center, we include it here. Throughout the paper, we set $M_{0}=1$ and $r_0=10$.
\end{itemize}

\subsection{Conditions on the parameters\label{sec:conditions}}
No stable circular orbits exist for $r\in[r_b;~r_{\scriptsize\mbox{ISCO}}]$ if the outer boundary of the sphere is inside the innermost stable circulate orbit: $r_b<r_{\scriptsize\mbox{ISCO}}=6M_{T}$~\cite{RN_1304}. So, for simplicity, we only consider the cases of $r_b\ge6M_{T}$. For the JMN2 model, in order to make sure that the density is positive and $r_b\ge6M_{T}$, there is $1/\sqrt{2}\le\lambda<1$. See Eq.~(\ref{JMN2_parameters}).

As implied by Eq.~(\ref{p_r}), inside the sphere, in order to avoid a pole in $p_{,r}$ and make sure that $p_{,r}<0$, there should be $F(r)<r$ for $r\le r_{b}$. In addition, this condition is also necessary to avoid apparent horizon formation.

\begin{figure*}[t!]
  \epsfig{file=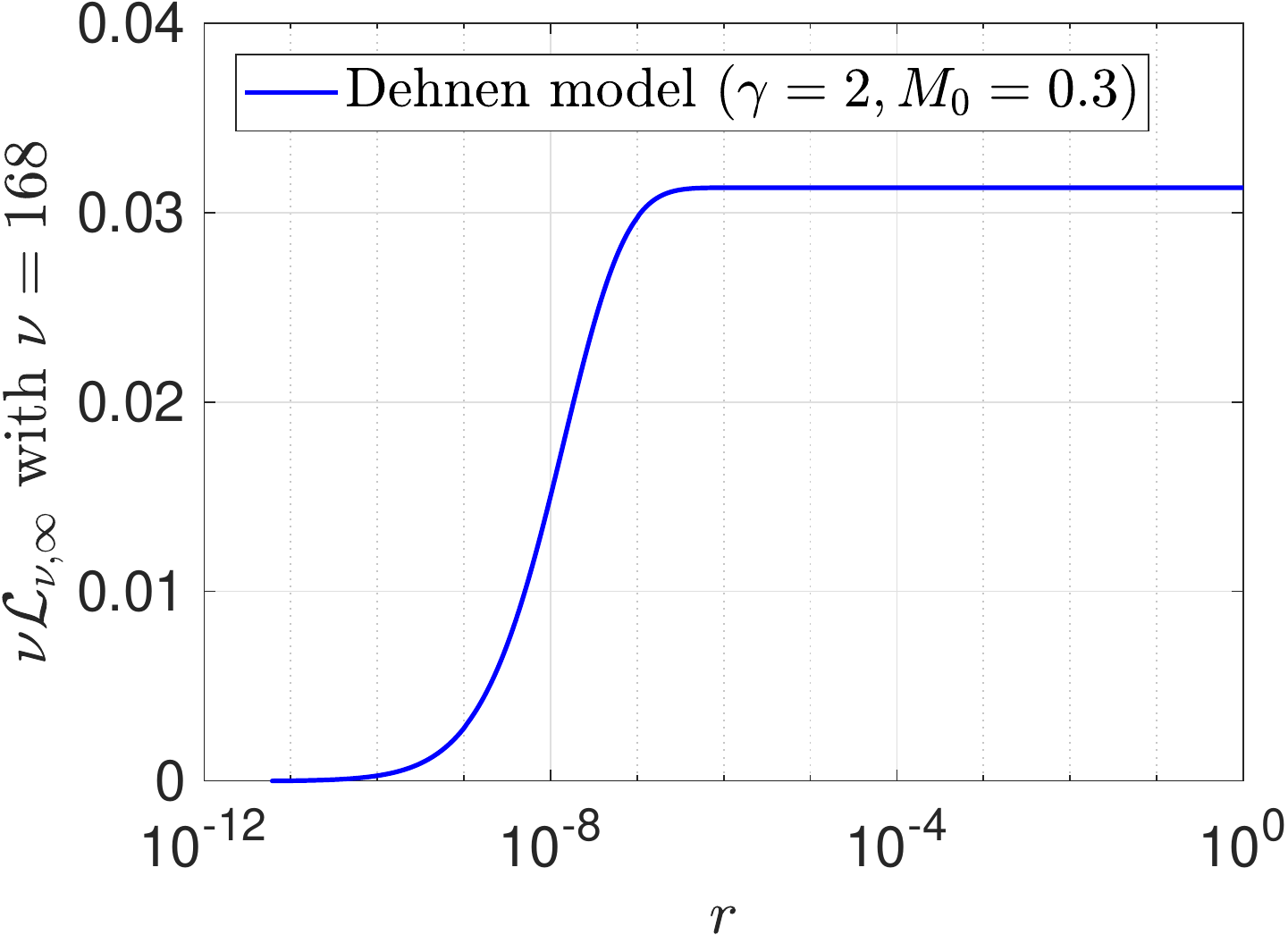,width=6cm}
  \caption{Contribution to the spectral luminosity distribution $\nu{\cal L}_{\nu,\infty}$ (\ref{integral_L}) from the radiation emitted at different radii for the Dehnen model with $\gamma=2$ and $M_{0}=0.3$. We use $\nu=168$. Correspondingly, $\log_{10}(h\nu/kT_{*})=2.5$ (see Figs.~\ref{fig:spectra} and \ref{fig:slope}). The results show that the luminosity distribution at high frequencies is mainly from the radiation emitted near the center of the galaxy.}
  \label{fig:luminosity_radius}
\end{figure*}

\subsection{Numerics}
On the outer boundary of the sphere $r=r_{b}$, which can be obtained from Eq.~(\ref{boundary_condition}), the pressure $p$ is zero, $e^{2\phi}=D=1-2M_{T}/r_b$, and $F(r_b)=2M_{T}$. With such quantities at $r=r_b$, we numerically integrate Eqs.~(\ref{phi_r}) and (\ref{p_r}) from $r=r_b$ toward $r=0$ via the fourth-order Runge-Kutta method, and respectively obtain $\phi(r)$ and $p(r)$. Outside the sphere, we have a Schwarzschild exterior spacetime. In this paper, we set $\dot{m}=M_{T}=1$.

In Eq.~(\ref{integral_L}), the radiation at different radii contributes differently to the spectral luminosity distribution. For the naked singularity models, the lower and upper limits of the radius in the integral in Eq.~(\ref{integral_L}) are $r=10^{-13}$ and $r=10^5$, respectively. This setup allows that the main contributions along the radii can always be included. As a sample check, the contributions to the spectral luminosity distribution along the radii for the Dehnen model with $\gamma=2$ at $\nu=168$ are plotted in Fig.~\ref{fig:luminosity_radius}.

\subsection{Code validation\label{sec:validation}}
At places far away from the matter cloud, the gravitational field is very weak, and can be well described by Newtonian gravity. The properties of the steady thin accretion disks were studied in Newtonian gravity in Refs.~\cite{FKR,Ryden}. Denote $T_{o}$ as the temperatures on the inner and outer boundaries of the disk. At very low frequencies, $\nu\ll kT_{o}/h$, the spectrum is mainly contributed by the Rayleigh-Jeans tail of the cool, outer edge of the accretion disk, and there is ${\cal L}_{\nu,\infty}\propto\nu^{2}$. At intermediate frequencies, $kT_{o}/h \ll \nu \ll kT_{*}/h$, there is ${\cal L}_{\nu,\infty}\propto\nu^{1/3}$. At high frequencies, $\nu\gg kT_{*}/h$, the integral for the spectrum is dominated by the hottest parts of the disk, and the spectrum is exponential. The numerical results for ${\nu}{\cal L}_{\nu,\infty}$ at low and intermediate frequencies, generated by the code and shown in Fig.~\ref{fig:slope}, are consistent with such analytical expressions. We also compute the spectrum at high frequencies for the Schwarzschild black hole model. The results match well with those in Newtonian gravity depicted in Refs.~\cite{FKR,Ryden}. See Fig.~\ref{fig:Schw}.

As described by Eq.~(\ref{efficiency}), the efficiency for the conversion of energy of the accreting gas into radiation can be obtained by integrating the luminosity distribution over frequency. Alternatively, one can obtain the efficiency from the energy loss for a test particle moving from infinity to the inner boundary of the disk. The energy per unit mass $E$ for a particle is expressed by Eq.~(\ref{E_accretion}). We denote the energies at $r=\infty$ and at the inner boundary of the radius as $E_{\infty}$ and $E_{\rm inner}$, respectively. Considering that at $r=\infty$, $E\approx1$, we arrive at
\be \mbox{Efficiency}=\frac{E_{\infty}-E_{\rm inner}}{E_{\infty}}\approx1-E_{\rm inner}.\label{efficiency_energy_loss}\ee
We compute the efficiencies by both the energy-loss method and luminosity-integration method. The results are well compatible. See Table~\ref{table:efficiency}. This is another validation of the code.

Note that the efficiencies for the density profile models are different. However, one cannot use such differences to distinguish the models since an independent estimate of the mass accretion rate $\dot{m}$ cannot be obtained.

\begin{table*}
    \begin{tabular}{ l | p{2.15cm} | p{2.15cm} }
      \hline
      \textbf{Model} & \textbf{Efficiency via luminosity integration} & \textbf{Efficiency via energy loss}  \\ \hline
      Schw. BH               & 0.05692 & 0.05719 \\ \hline
      JMN2 model ($\lambda=1/\sqrt{2}$)  & 1.0015  & 0.9999  \\ \hline
      Dehnen model ($\gamma=1, M_{0}=1$)     & 0.3491  & 0.3493  \\ \hline
      Dehnen model ($\gamma=1.5, M_{0}=0.4$) & 0.35130 & 0.35132 \\ \hline
      Dehnen model ($\gamma=2, M_{0}=0.3$)   & 1.0006  & 0.9986  \\ \hline
      NFW model   ($\gamma=2, M_{0}=1$)      & 0.2554  & 0.2557  \\
    \hline
    \end{tabular}
   \\[10pt]
  \caption{Efficiencies for the accretion disks by the luminosity-integration method~(\ref{efficiency}) and energy-loss method~(\ref{efficiency_energy_loss}).}
\label{table:efficiency}
\end{table*}

\begin{figure*}[t!]
  \epsfig{file=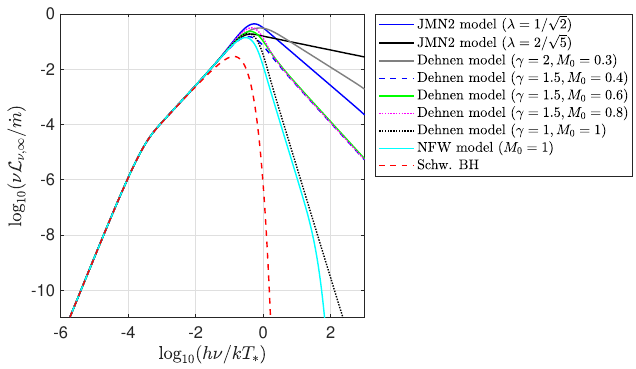, width=0.6\textwidth}
  \caption{(color online). Spectral luminosity distributions of radiation from the accretion disks for the density profile models. Note that the results for the NFW model for $\log_{10}(h\nu/kT_{*})>1$ are not accurate.}
  \label{fig:spectra}
\end{figure*}

\section{Result I: Thermal properties of the accretion disks\label{sec:luminosity}}
\begin{figure*}[t!]
  \epsfig{file=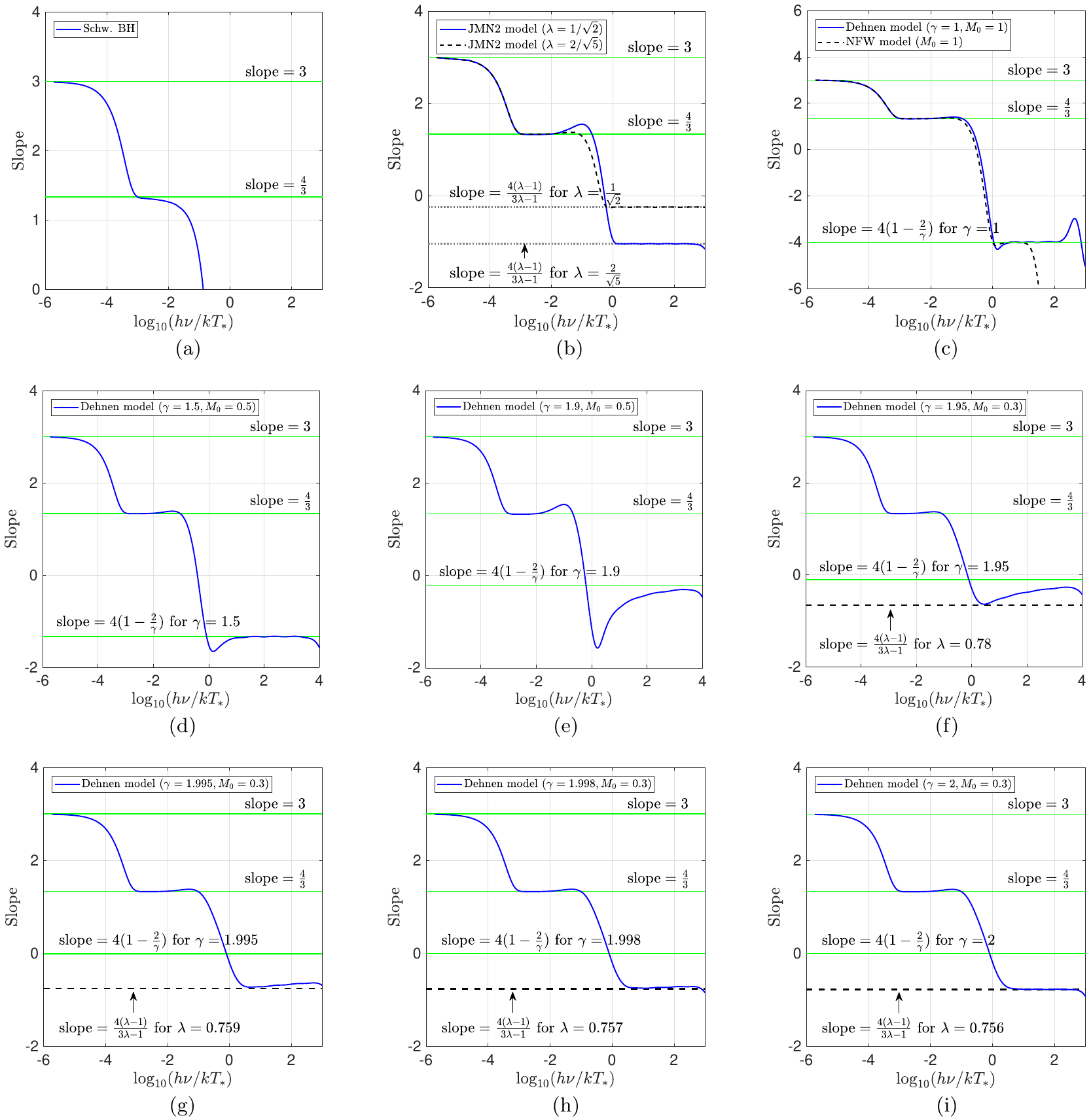,width=0.9\textwidth}
  \caption{Slopes of the spectral luminosity distributions for the accretion disks in the density profile models,
  $\mbox{slope}{\equiv}d[\log_{10}(\nu {\cal L}_{\nu,\infty}/\dot{m})]/d[\log_{10}(h\nu/kT_{*})]$.
  (a)-(i): At very low frequencies where $\nu\ll kT_{o}/h$, there is ${\nu}{\cal L}_{\nu,\infty}\propto\nu^{3}$. At intermediate frequencies where $kT_{o}/h \ll \nu \ll kT_{*}/h$, there is ${\nu}{\cal L}_{\nu,\infty}\propto\nu^{4/3}$.
  (c)-(e): For the Dehnen model with $1\le\gamma<2$, at high frequencies, there is $\nu {\cal L}_{\nu,\infty}\propto\nu^{4(1-2/\gamma)}$.
  (f)-(h): For the Dehnen model with $\gamma$ close to 2, for certain range of small radii where $r^{2-\gamma}\to 1$, there is $\phi\approx\phi_0+{M_0}/(2\zeta)+M_{0}\ln r/2$. This is close to the JMN2 model. Correspondingly, for certain range of high frequencies, $\nu{\cal L}_{\nu,\infty} \propto\nu^{4(\lambda-1)/(3\lambda-1)}$, with $\lambda=\sqrt{(1-2M_{0}^{\scriptsize\mbox{new}})/(1-M_{0}^{\scriptsize \mbox{new}})}$ and $M_{0}^{\mbox{new}}\approx(3-\gamma)M_0/r_0^{2-\gamma}\approx{M_0}$. As $r$ is even closer to zero, $r^{2-\gamma}\ll(2-\gamma)\ll 1$. Then the expression for $\phi$ comes back to $\phi\approx\phi_0$. Correspondingly, at even higher frequencies, the expression for
  $\nu {\cal L}_{\nu,\infty}$ comes back to $\nu {\cal L}_{\nu,\infty}\propto\nu^{4(1-2/\gamma)}$.
  (i): When $\gamma=2$, at high frequencies, there is always $\nu{\cal L}_{\nu,\infty} \propto\nu^{4(\lambda-1)/(3\lambda-1)}$. Note that the numerical results for $\log_{10}(h\nu/kT_*)>3$ are not accurate.}
  \label{fig:slope}
\end{figure*}

In this section, we investigate the accretion disks' thermal properties, including the spectral luminosity distribution and efficiencies. The spectra at low frequencies are mainly determined by the accreting matter at large radii, and those at high frequencies by the accreting matter at small radii. As discussed in Sec.~\ref{sec:framework} and shown in Fig.~\ref{fig:slope}, at low frequencies, the spectra for all the models have the same shape. In Subsection~\ref{expression_high_frequency}, we explore those at high frequencies, and find that they have dramatically different shapes, which can be used to distinguish the density profile models and the Schwarzschild black hole model.

In addition, we will interpret why the efficiencies for some density profile models do not approach $1$. The transition of the Dehnen model with $\gamma\approx 2$ to the JMN2 model will be discussed.

\subsection{Spectral luminosity distribution}
Numerically integrating Eq.~(\ref{integral_L}), we obtain the spectral luminosity distributions for the accretion disks, with the results being plotted in Fig.~\ref{fig:spectra}. For a given mass accretion rate $\dot{m}$, at low frequencies, all the models have the same shape. This is the standard result for disk emission from large non-relativistic radii.

At high frequencies, the luminosities for the JMN2 model and Dehnen model with $\gamma=2$ are greater than those for the Dehnen model with $\gamma<2$, the NFW model and the Schwarzschild black hole model. As discussed in Sec.~\ref{sec:validation}, this is not observationally testable since $\dot{m}$ cannot be estimated independently. As will be investigated in the next subsection, more interestingly, the spectra have noticeably different shapes at high frequencies, which can in theory be used to distinguish the density profile models and the Schwarzschild black hole model.

\subsection{Asymptotic analytic expressions for the spectra at high frequencies\label{expression_high_frequency}}

\subsubsection{JMN2 model}
For the JMN2 model, with the information presented in Sec.~\ref{sec:framework}, one can obtain the asymptotic expressions for the metric component $A(=g_{tt})$, energy per unit mass $E$, angular momentum per unit mass $L$, angular velocity $\omega$, redshift $z$, determinant of the metric $g$ and so on near the center:
\be A\approx \alpha^{2}r^{2(1-\lambda)},\ee
\be E\sim A^{1/2}\propto r^{1-\lambda},\label{E_JMN2}\ee
\be L\propto r,\label{L_JMN2}\ee
\be \omega\propto r^{-\lambda},\ee
\be 1+z(r)=[-(g_{tt}+\omega^2 g_{\varphi\varphi})]^{-1/2}\propto r^{-1+\lambda},\ee
\be \sqrt{-g}=\sqrt{2-\lambda^2}r A^{1/2}\propto r^{2-\lambda},\ee
\be \int^{r}_{0}(E-\omega L)L_{,{\tilde{r}}}d{\tilde{r}}\propto r^{2-\lambda},\ee
\be
\mathcal{F}(r)=-\frac{\dot{m}}{4\pi \sqrt{-g}}\frac{\omega_{,r}}
{(E-\omega L)^2}\int^r_{0}(E-\omega L)L_{,{\tilde{r}}}d{\tilde{r}}
\propto r^{-3+\lambda},
\ee
\be \frac{d{\cal L}_\infty}{d\ln r}=4\pi r\sqrt{-g} E {\mathcal F}\propto r^{2\lambda}.\ee
With the above expressions and Eq.~(\ref{integral_L}), the luminosity distribution at high frequencies observed at infinity can be obtained~\cite{RN_1304},
\beq
\nu {\cal L}_{\nu,\infty} &=& \frac{15}{\pi^4} \int_{0}^\infty \left(\frac{d{\cal L}_\infty}{d\ln r}\right)
\frac{(1+z)^4 (h\nu/kT_*)^4/\mathcal{F}}{\exp[(1+z)(h\nu/kT_*)/{\mathcal F}^{1/4}]-1}\, d\ln r\nonumber\\
&\sim& \nu^{4}\int_{0}^{r_{2}}\frac{1}{e^{x}-1}r^{2\lambda-1}\, dr\nonumber\\
&\sim& \nu^{4(\lambda-1)/(3\lambda-1)}\int_{0}^{x_{2}}\frac{x^{4(2\lambda-1)/(3\lambda-1)}}{e^{x}-1}\, dx\nonumber\\
&\propto& \nu^{4(\lambda-1)/(3\lambda-1)},
\label{approx_JMN2}
\eeq
where $x=(1+z)[h\nu/(kT_{*})]/{\mathcal F}^{1/4}\propto\nu r^{(3\lambda-1)/4}$ and $r_{2}\ll 1$.

We define the slope $J$ of the spectra with respect to the frequency as
\be J{\equiv}\frac{d[\log_{10}(\nu {\cal L}_{\nu,\infty}/\dot{m})]}{d[\log_{10}(h\nu/kT_{*})]}.\label{slope}\ee
Then with Eqs.~(\ref{approx_JMN2}) and (\ref{slope}), the slope for the JMN2 model at high frequencies takes the form
\be J_{\scriptsize \mbox{JMN2}}\approx\frac{4(\lambda-1)}{3\lambda-1},\label{slope_JMN2}\ee
which is verified by the numerical results plotted in Fig.~\ref{fig:slope}(b).

\subsubsection{Dehnen model}
For the Dehnen model, the numerical results show that near the center $F/(pr^3)\gg1$. This leads to
\be \phi_{,r}=\frac{pr^3+F}{2r(r-F)}\approx\frac{F}{2r^2}\approx\frac{M_0}{2r_{0}^{2-\gamma}}r^{1-\gamma}.\label{dphidr_Dehnen}\ee
For $1\le\gamma<2$ and $r^{2-\gamma}/(2-\gamma)\to 0$, there are
\be \phi-\phi_0\approx\frac{M_0}{2r_{0}^{2-\gamma}(2-\gamma)}r^{2-\gamma}\to 0,\label{phi_Dehnen}\ee
\be A\approx e^{2\phi_0}\exp\left[\frac{M_{0}}{2-\gamma}\left(\frac{r}{r_{0}}\right)^{2-\gamma}\right]\approx e^{2\phi_0},\ee
\be E\approx A^{1/2}\approx e^{\phi_0},\label{E_Dehnen}\ee
\be L\propto r^{2-\gamma/2},\label{L_Dehnen}\ee
\be \omega\propto r^{-\gamma/2},\ee
\be 1+z(r) = [-(g_{tt}+\omega^2 g_{\varphi\varphi})]^{-1/2} \sim e^{\phi_0},\ee
\be \sqrt{-g}\propto r,\ee
\be
\mathcal{F}(r)=-\frac{\dot{m}}{4\pi \sqrt{-g}}\frac{\omega_{,r}}
{(E-\omega L)^2}\int^r_{0}(E-\omega L)L_{,\tilde{r}}d\tilde{r}
\propto r^{-\gamma},
\ee
\be \frac{d{\cal L}_\infty}{d\ln r}=4\pi r\sqrt{-g} E {\mathcal F}\propto r^{2-\gamma}.\ee
Then the luminosity distribution at high frequencies is
\beq
\nu {\cal L}_{\nu,\infty} &=& \frac{15}{\pi^4} \int_{0}^\infty \left(\frac{d{\cal L}_\infty}{d\ln r}\right)
\frac{(1+z)^4 (h\nu/kT_*)^4/\mathcal{F}}{\exp[(1+z)(h\nu/kT_*)/{\mathcal F}^{1/4}]-1}\, d\ln r\nonumber\\
&\sim& \nu^{4}\int_{0}^{r_{2}}\frac{1}{e^x-1}r\, dr\nonumber\\
&\sim& \nu^{4\left(1-2/\gamma\right)}\int_{0}^{x_{2}}\frac{x^{8/\gamma-1}}{e^x-1}\, dx\nonumber\\
&\propto& \nu^{4\left(1-2/\gamma\right)},
\label{approx_Dehnen}
\eeq
where $x=(1+z)(h\nu/kT_*)/{\mathcal F}^{1/4}\propto\nu r^{\gamma/4}$ and $r_{2}\ll 1$. Then using Eqs.~(\ref{slope}) and (\ref{approx_Dehnen}), the slope of the spectra for the Dehnen model at high frequencies takes the form
\be J_{\scriptsize \mbox{Dehnen}}\approx4\left(1-\frac{2}{\gamma}\right),\label{slope_Dehnen}\ee
which is confirmed by the numerical results plotted in Figs.~\ref{fig:slope}(c) and \ref{fig:slope}(d).

Near the center, the density profile for the NFW model is close to that for the Dehnen model with $\gamma=1$. As expected, the luminosity distribution at high frequencies for this model is close to that for the Dehnen model with $\gamma=1$. See Fig.~\ref{fig:slope}(c). The results for $\gamma=2$ are the same as those in the JMN2 model. See Figs.~\ref{fig:slope}(b) and \ref{fig:slope}(i).

\begin{figure*}[t!]
  \epsfig{file=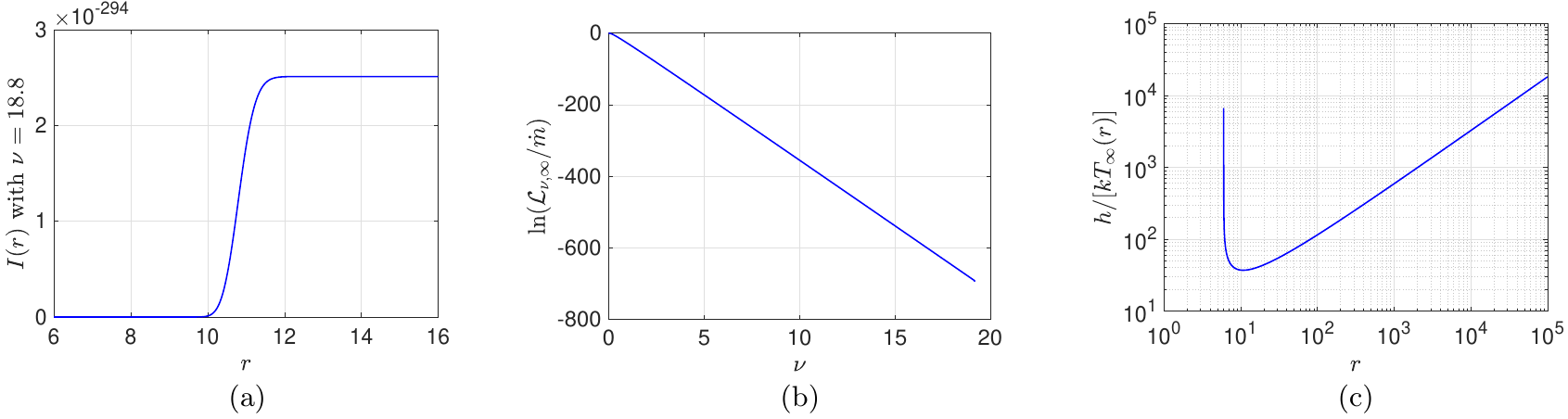,width=\textwidth}
  \caption{Spectral luminosity distribution at high frequencies $\nu$ for the accretion disk in the Schwarzschild black hole model. (a) $I(r)|_{\nu=18.8}$ vs. $r$. $I(r)$ is obtained with the upper limit $\infty$ of the integral (\ref{integral_L}) being replaced by $r$. (b) $\ln({\cal L}_{\nu,\infty}/\dot{m})$ vs. $\nu$. $\mbox{Slope}\approx-36.77$ for $18.5<\nu<19.0$. (c) $h/[kT_{\infty}(r)]$ vs. $r$. The minimum for the $h/[kT_{\infty}(r)]$ is about $36.90$ at $r\approx10.79$.}
  \label{fig:Schw}
\end{figure*}

\subsubsection{The Schwarzschild black hole model}
In Newtonian gravity, at high frequencies, the integral for the spectrum is mainly contributed by the hottest parts of the disk. Then the spectrum is exponential~\cite{FKR,Ryden}. We also compute the spectrum at high frequencies for the Schwarzschild black hole model. The results match well with those in the Newtonian gravity case depicted in Refs.~\cite{FKR,Ryden}. As shown in Fig.~\ref{fig:Schw}(b), at high frequencies,
\be \ln\left(\frac{{\cal L}_{\nu,\infty}}{\dot{m}}\right)\propto-\nu.\label{slope1_Schw}\ee
The slope for the plot of $\ln({\cal L}_{\nu,\infty}/\dot{m})$ vs. $\nu$ at $18.5<\nu<19.0$ is about $-36.77$. Note that the exponential function in Eq.~(\ref{integral_L}) is $e^{-h\nu/[kT_{\infty}(r)]}$. As shown in Fig.~\ref{fig:Schw}(c), the minimum value for $h/[kT_{\infty}(r)]$ is about $36.90$ at $r\approx10.79$. So the slope of the spectra for the Schwarzschild black hole model at high frequencies is
\be J_{\scriptsize \mbox{Schw}}\approx-\frac{h\nu}{kT_{(\infty,\mbox{\scriptsize min})}},\label{slope2_Schw}\ee
where $T_{(\infty,\mbox{\scriptsize min})}$ is the minimum of $T_{\infty}(r)$.

The slope of the spectral luminosity distribution does not depend on the mass accretion rate $\dot{m}$. On the other hand, as shown by Eqs.~(\ref{approx_JMN2}), (\ref{approx_Dehnen}) and (\ref{slope2_Schw}) and Figs.~\ref{fig:slope} and \ref{fig:Schw}, the slopes at high frequencies for the models are different. So this feature is observationally testable and can be used to distinguish the models, especially to distinguish the naked singularity models from the Schwarzschild black hole model.

\subsubsection{Transition between the Dehnen with $\gamma\approx2$ and JMN2 models}
When $\zeta(\equiv2-\gamma)$ and $r$ are close to zero, the expression for $\phi$~(\ref{phi_Dehnen}) becomes subtle and deserves further explorations. When $r^\zeta\to1$, namely
$|\zeta\ln r|\ll1$, there is
\be r^\zeta=e^{\zeta\ln r}\approx1+\zeta\ln r.\ee
Then Eq.~(\ref{phi_Dehnen}) can be rewritten as
\be \phi\approx\phi_0+\frac{M_0}{2\zeta}+\frac{M_0}{2}\ln r,\label{phi_sln_log}\ee
which is close to that in the JMN2 model. See Eq.~(\ref{A_JMN2}). Correspondingly, at certain range of high frequencies, $\nu{\cal L}_{\nu,\infty}$ takes the expression~(\ref{approx_JMN2}) as in the JMN2 model.

As $r$ is even closer to zero, there will be $r^\zeta\ll\zeta\ll 1$. Then as in the normal case of $\gamma<2$, the expression for $\phi$ ~(\ref{phi_Dehnen}) will transit back to
the normal case of the Dehnen model,
\be \phi\approx\phi_0.\ee
Consequently, at higher frequencies, $\nu{\cal L}_{\nu,\infty}$ takes the expression~(\ref{approx_Dehnen}). This is partially confirmed in Figs.~\ref{fig:slope}(f)-\ref{fig:slope}(i). In Fig.~\ref{fig:slope}(f) for which $\zeta=0.05$, the slope for the luminosity distribution touches the line of $\mbox{slope}=4(\lambda-1)/(3\lambda-1)$ around $\log_{10}(h\nu/kT_{*})\approx0.4$, then moves toward the line of $\mbox{slope}=4(1-2/\gamma)$ as $\nu$ increases, and almost touches the line of $\mbox{slope}=4(1-2/\gamma)$ around $\log_{10}(h\nu/kT_{*})\approx3.6$ where the numerical computations break down. We expect the slope to touch and stay on the line of $\mbox{slope}=4(1-2/\gamma)$ at higher frequencies.

\begin{figure*}[t!]
  \epsfig{file=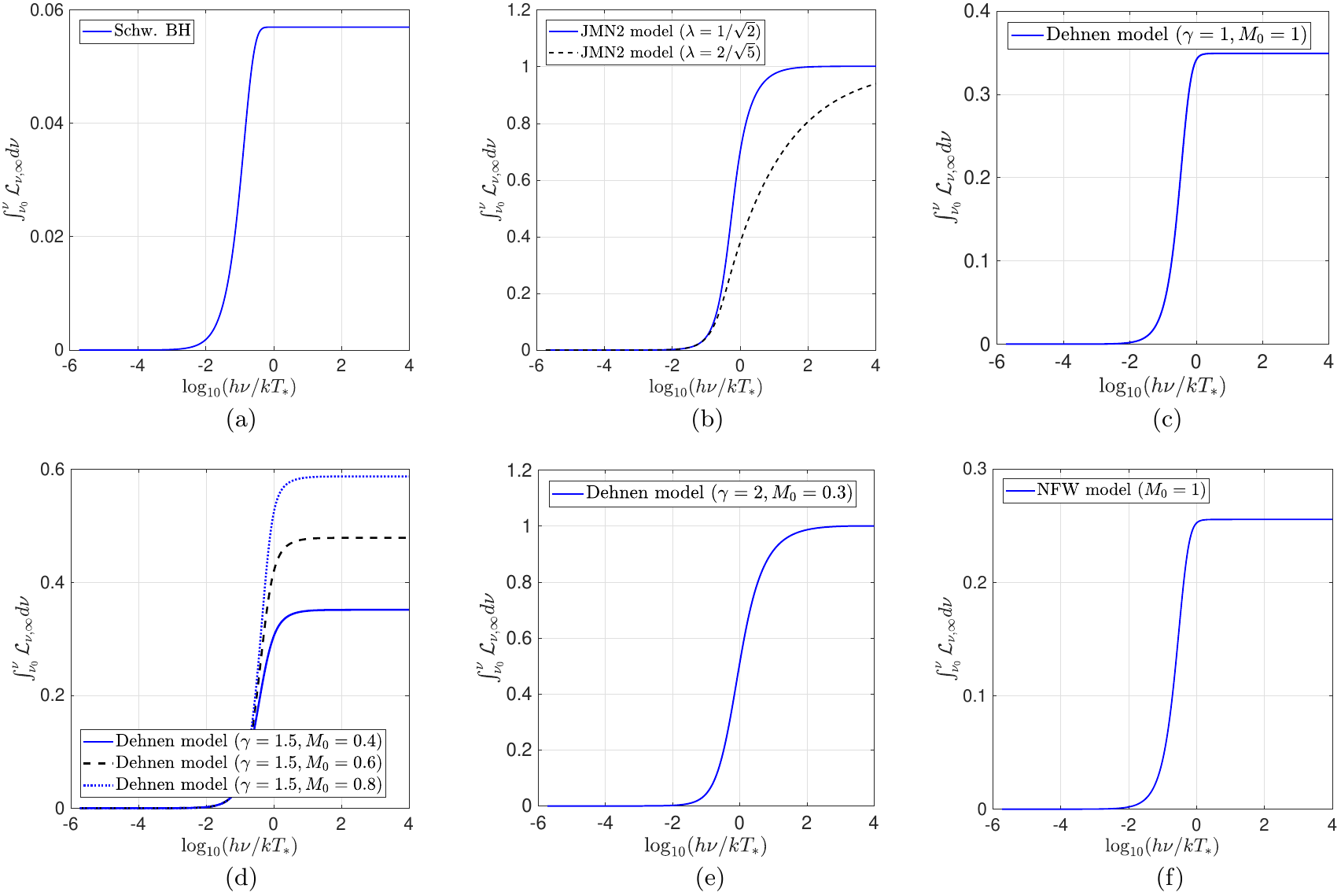,width=0.9\textwidth}
  \caption{$\int_{\nu_0}^{\nu}{\cal L}_{\nu,\infty}d\nu$ vs. $\nu$ for some density profile models. $\nu_{0}=10^{-6}$. According to Eq.(\ref{efficiency}),
  $\mbox{Efficiency}=\int_{0}^{\infty}{\cal L}_{\nu,\infty}d\nu$. As show in the figure, the plots of $\int_{\nu_0}^{\nu}{\cal L}_{\nu,\infty}d\nu$ vs. $\nu$ are very flat at very low and very high frequencies, implying that the main contribution of ${\cal L}_{\nu,\infty}$ along the frequencies has been included. For the JMN2 model in (b) and the Dehnen model with $\gamma=2$ in (e), the efficiencies are equal to $1$; while for the Dehnen model with $\gamma<2$ in (c) and (d) and the NFW model in (f), the efficiencies do not approach $1$.}
  \label{fig:efficiency}
\end{figure*}

\begin{figure*}[t!]
  \epsfig{file=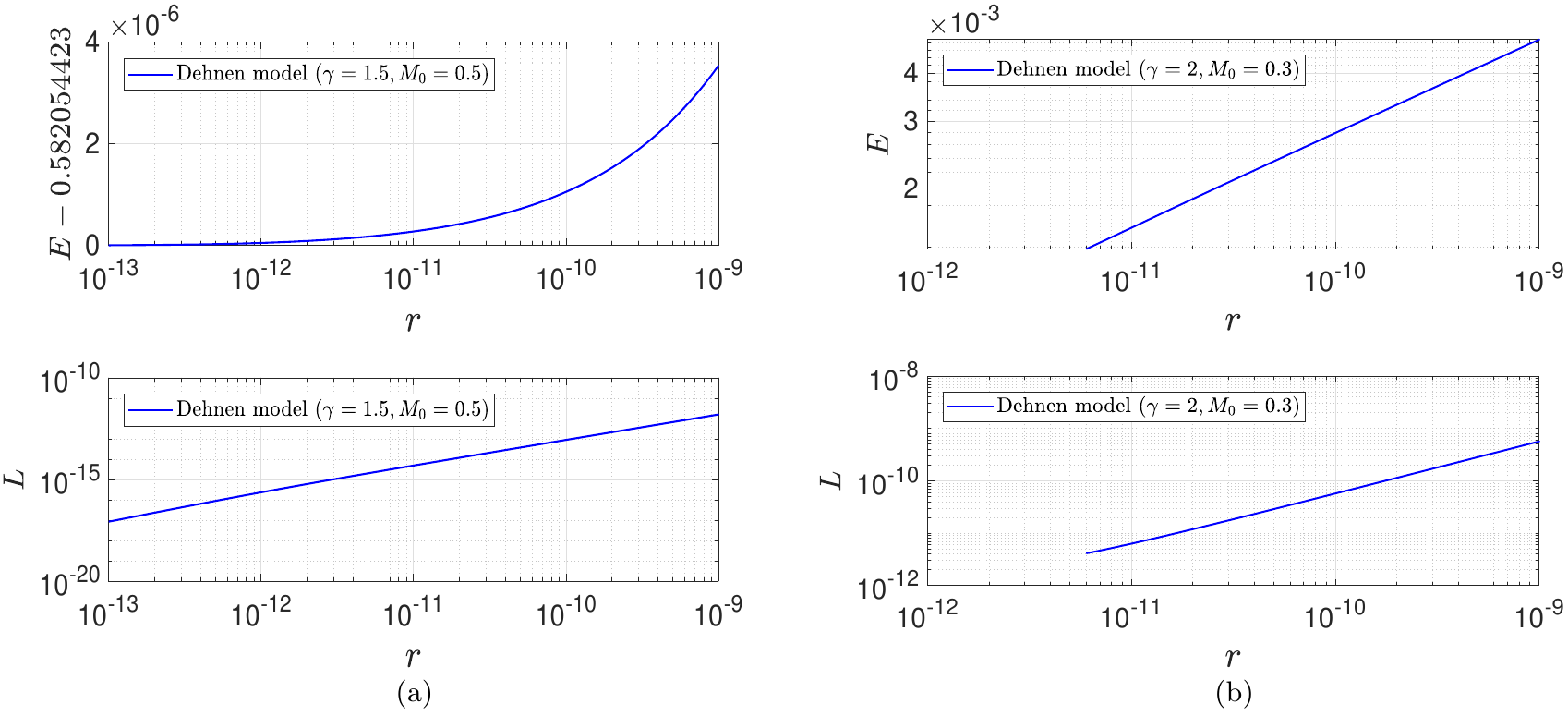,width=0.8\textwidth}
  \caption{Energy per unit mass $E$ and angular momentum per unit mass $L$ for the Dehnen model near the center. (a) When $\gamma=1.5$, although $L$ goes to zero as the center is approached, $E$ does not. (b) When $\gamma=2$, both $E$ and $L$ go to zero as the center is approached.}
  \label{fig:E_L}
\end{figure*}

\begin{figure*}[t!]
  \epsfig{file=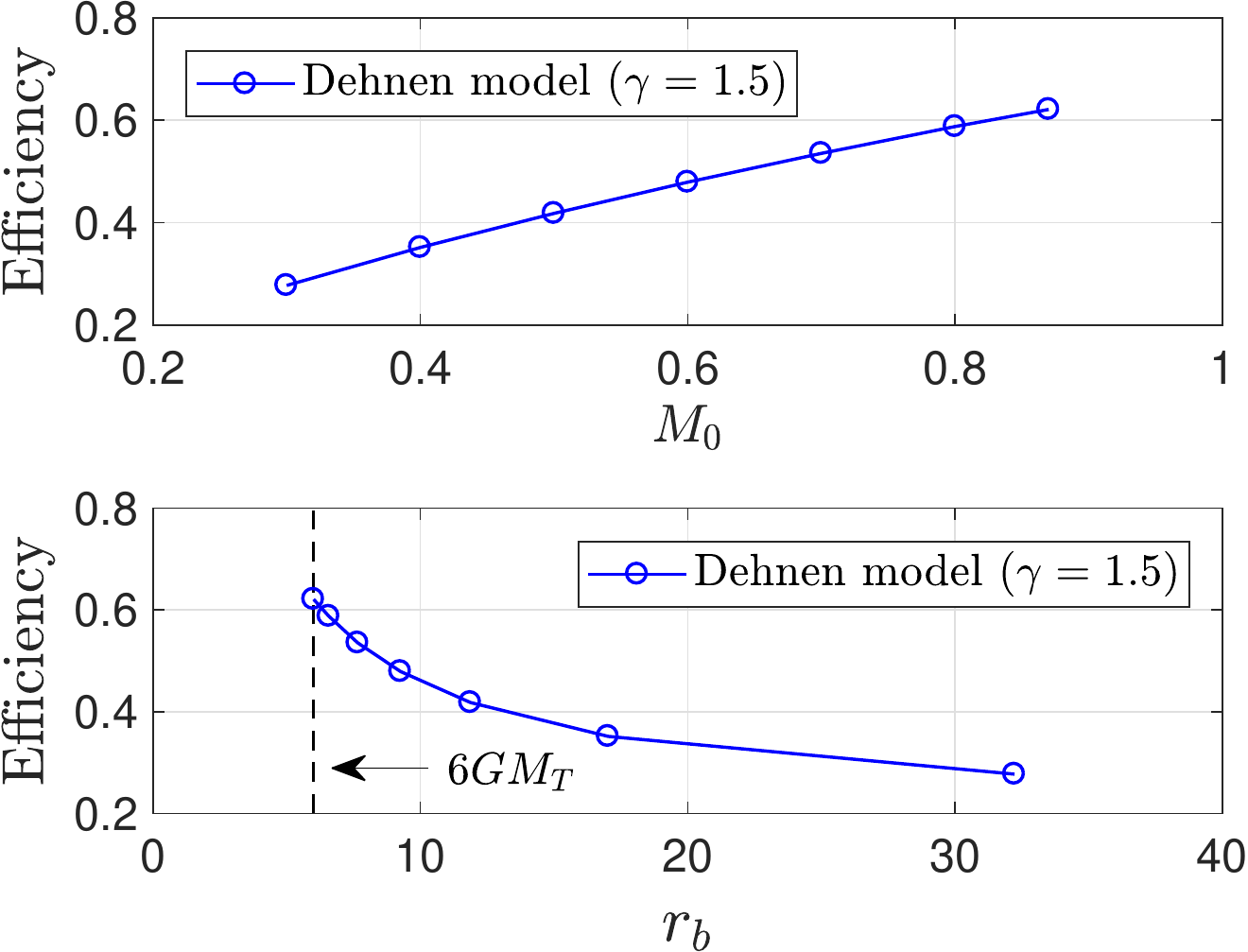,width=6.23cm}
  \caption{Efficiency vs. $M_{0}$ and efficiency vs. $r_b$ for the Dehnen model with $\gamma=1.5$. The more compact the sphere is, the higher is the efficiency. The constraint of $r_b\ge6M_{T}$ makes the efficiency not approach $1$.}
  \label{fig:efficiency_vs_compactness}
\end{figure*}

\subsubsection{Feasibility of distinguishing models through observations}
A feature of all the density profile models discussed in this paper is
the presence of a power-law tail in the accretion disk spectrum at
high frequencies (see Fig.~\ref{fig:spectra}). This tail is quite
distinct from the exponential cutoff in the disk spectrum for a
Schwarzschild black hole. At first sight it appears that this provides
an easy way of distinguishing the two kinds of model. However, there
are practical difficulties because accreting black holes in Nature
rarely follow precise model predictions. In particular, all accreting
black holes show power-law tails in their spectra, believed to be
produced by Compton scattering in hot coronal gas above the disk
proper (see \cite{mcclintock_2006} for observed spectra of
accreting stellar-mass black holes in X-ray binaries).

Among the many spectral states observed in black holes binaries
(BHBs), of particular interest is the thermal-dominant or high/soft
state, which comes closest to matching the predictions of thin
accretion disk models.  X-ray spectra in this state show a peak with a
sharp cutoff, just like the model Schwarzschild black hole spectrum in
Fig.~\ref{fig:spectra}, followed by a weak power-law tail at higher
frequencies. Example spectra are shown in Fig.~4.8 of
\cite{mcclintock_2006} and are qualitatively quite different from the
spectra computed in the present paper for the JMN2 and Dehnen
models. Specifically, in our spectra the power-law tail originates
directly from the peak of the spectrum, whereas the thermal-dominant
state spectra of BHBs have a well-defined cutoff beyond the peak, and
the faint power-law component emerges only after the luminosity has
fallen substantially below its peak value. The presence of a
thermal-dominant spectrum in a BHB would thus rule out the models
described in the present paper. Conversely, if a BHB in the
appropriate luminosity range ($\sim 5-50\%$ of Eddington) does not
exhibit a thermal-dominant spectrum, it would be a candidate for one
of our models.

\subsection{Efficiencies for the accretion disks}
Integrating Eq.~(\ref{efficiency}), we obtain the efficiencies for the accretion disks. As shown in Fig.~\ref{fig:efficiency}, except the JMN2 model and the Dehnen model with $\gamma=2$ for which the efficiencies are equal to $1$, the efficiencies for all other models are less than $1$. Note that, as shown in Eq.~(\ref{E_JMN2}), for the JMN2 model and the Dehnen model with $\gamma=2$, the energy per unit mass for a test particle goes to zero as the center is approached. Also see Fig.~\ref{fig:E_L}(b) for numerical results. So all the mass energy of the accreting gas is converted into radiation: the efficiency is $1$. For $1\le\gamma<2$, as shown in Eqs.~(\ref{E_Dehnen}) and (\ref{L_Dehnen}), as the center is approached, the angular momentum per unit mass $L$ goes to zero, while the energy per unit mass $E$ does not. This is also confirmed by the numerical results in Fig.~\ref{fig:E_L}(a). Therefore, not all the mass energy of the accreting gas is converted into radiation: the efficiency is less than $1$.

Take the case of $\gamma=1.5$ as an example, we find that the efficiency for the Dehnen model increases when the parameter $M_{0}$ increases. See Figs.~\ref{fig:efficiency}(d) and \ref{fig:efficiency_vs_compactness}. Alternatively, the efficiency increases when the sphere becomes more compact (the radius of the sphere $r_b$ decreases). Is it possible for the sphere to be compact enough such that the efficiency goes to $1$? As mentioned in Sec.~\ref{sec:conditions}, no stable circular orbits exist for $r\in[r_b;~r_{\scriptsize\mbox{ISCO}}]$ if $r_b<r_{\scriptsize\mbox{ISCO}}=6M_{T}$. Under the constraint of $r_b\ge6M_{T}$, the efficiency cannot approach $1$.
See Fig.~\ref{fig:efficiency_vs_compactness}.

\section{Result II: strength of the naked singularities\label{sec:strength}}
In this section, we study the strength of the naked singularities in some density profile models. According to the Tipler's definition~\cite{Tipler_1977}, a singularity is called gravitationally strong if the volume (area) element defined by three (two) independent vorticity-free Jacobi fields along a timelike (null) geodesic toward the singularity goes to zero as the singularity is approached,
\be \displaystyle{\lim_{\tau \to 0}}V(\tau)=0,\ee
where $\tau$ is the proper time. For convenience, in this paper, we set $\tau=0$ at the singularity, and positive before the singularity is approached. Based on the condition by Clarke and Krolak~\cite{Clarke_1985}, a singularity is strong in the sense of Tipler if along the non-spacelike geodesic, as the singularity is approached, there is
\be \displaystyle{\lim_{\tau \to 0}}\tau^{2}R_{ab}T^{a}T^{b}>0,\label{condition_strong}\ee
where $R_{ab}$ is the Ricci tensor, and $T^{a}(\equiv dx^{a}/d\tau)$ a tangent vector to the geodesic.

In this section, we first present the formalism for computing $V(\tau)$ and $\tau^2 R_{ab}T^{a}T^{b}$. Then we analyse the strength of the singularities in the JMN2 and Dehnen models. After that, the Kretschmann scalar near the singularities in the two models will be studied, and the connection between the efficiencies for the accretion disks and strength of the singularities will be discussed.

\subsection{Formalism}
Consider a radial timelike geodesic terminating at a naked central singularity. The $r$ and $t$ components of the geodesic equation are respectively
\be \ddot{r}+\frac{B'}{2B}\dot{r}^2+\frac{A'}{2B}\dot{t}^2=0,\label{geodesic_r_1}\ee
\be \ddot{t}+\frac{A'}{A}\dot{t}\dot{r}=\ddot{t}+\frac{\dot{A}}{A}\dot{t}=0,\label{geodesic_t}\ee
where the prime $(')$ and the dot $(^{.})$ denote derivatives with respect to the radius $r$ and proper time $\tau$, respectively. For the metric (\ref{metric}), $ds^2=-d\tau^2=-A{dt^2}+B{dr^2}+r^2 d\Omega^2$, there is
\be \dot{t}^2=\frac{B}{A}\dot{r}^2+\frac{1}{A}.\label{connection_t_r}\ee
Substitution of Eq.~(\ref{connection_t_r}) into (\ref{geodesic_r_1}) yields
\be \ddot{r}+\frac{1}{2}\left(\frac{A'}{A}+\frac{B'}{B}\right)\dot{r}^2+\frac{A'}{2AB}=0.\label{geodesic_r_2}\ee
Equation~(\ref{geodesic_t}) implies that
\be \dot{t}=\frac{C_1}{A},\label{geodesic_t_2}\ee
where $C_1$ is an integration constant.

The two Jacobi fields in the tangential 2-space can be written as
\be \overrightarrow{\xi}_{(1)}=x(t,r)\frac{\partial}{\partial\theta}, \hphantom{dddd}\overrightarrow{\xi}_{(2)}=y(t,r)\csc\theta\frac{\partial}{\partial\varphi}.\ee
Let one tangent vector to the radial timelike geodesic take the form
\be \overrightarrow{T}=\frac{d}{d\tau}=\dot{t}\frac{\partial}{\partial t}+\dot{r}\frac{\partial}{\partial r},\ee
then the Jacobi field in the radial 2-space orthogonal to $\overrightarrow{T}$ can be expressed as
\be \overrightarrow{\xi}_{(3)}=a(t,r)B\dot{r}\frac{\partial}{\partial t}+a(t,r)A\dot{t}\frac{\partial}{\partial r}.\ee
The geodesic deviation equation for $\overrightarrow{\xi}_{(1)}$ leads to
\be r\ddot{x}+2\dot{r}\dot{x}=0,\label{equation_x}\ee
which takes the same form as the one in the double-null coordinates~\cite{Nolan:1999tw}. Integrating Eq.~(\ref{equation_x}), we obtain
\be x(\tau)=C_{2}\int_{\tau_1}^\tau\frac{d\tau'}{r^{2}(\tau')},\label{solution_x}\ee
in which the initial condition, $x(\tau_1)=0$, has been set. The quantity $y(\tau)$ takes the same equation as $x(\tau)$ does. The geodesic deviation equation for $\overrightarrow{\xi}_{(3)}$ generates
\be
\ddot{a}+\frac{1}{2}\dot{a}\dot{r}\left(\frac{A'}{A}+\frac{B'}{B}\right)+\frac{1}{2}a\left[\dot{r}^2\left(-\frac{A'^2}{A^2}-\frac{B'^2}{B^2}+\frac{A''}{A}+\frac{B''}{B}\right)
+\frac{1}{B}\left(-\frac{A'^2}{A^2}-\frac{A'B'}{AB}+\frac{A''}{A}\right)\right]=0.
\label{equation_a}
\ee
For simplicity, we set $x=y$. Then the norm of the volume element defined by the three Jacobi fields is~\cite{Nolan:1999tw}
\be \parallel V(\tau)\parallel=|a|x^{2}r^2.\label{volume}\ee
The derivations of Eqs.~(\ref{equation_x}) and (\ref{equation_a}) are given in Appendix~\ref{sec:appendix_Jacobi_eq}.

Regarding the condition~(\ref{condition_strong}), the nonzero components of $R_{\alpha\beta}$ related to the problem we work on are
\be R_{tt}=-\frac{1}{4AB^{2}r}(ArA'B'-2ABrA''+A'^{2}Br-4ABA'),\label{Rtt}\ee
\be R_{rr}=\frac{1}{4A^2Br}(ArA'B'-2ABrA''+A'^{2}Br+4A^2B').\label{Rrr}\ee
Then the condition~(\ref{condition_strong}) becomes
\be \displaystyle{\lim_{\tau \to 0}}\tau^2 R_{ab}T^{a}T^{b}=\displaystyle{\lim_{\tau \to 0}} \tau^2 (R_{tt}\dot{t}^2+R_{rr}\dot{r}^2)>0.\label{condition_strong_2}\ee

\subsection{JMN2 model}
For the JMN2 model, $B=1/(2-\lambda^2)=\mbox{Const}$. Near the center, $A\approx\alpha^{2}r^{2(1-\lambda)}$. Then the geodesic equation (\ref{geodesic_r_2}) can be simplified as
\be \ddot{r}\approx-\frac{(1-\lambda)}{B}\frac{1+B\dot{r}^2}{r}.\label{geodesic_r_JMN2_approx}\ee
Multiply both sides of Eq.~(\ref{geodesic_r_JMN2_approx}) by $2B\dot{r}/(1+B\dot{r}^2)$, integrate, and note that $\tau=0$ at the center, one obtains
\be \dot{r}^2\approx\frac{1}{B}\left[C_{3}r^{-2(1-\lambda)}-1\right]\approx\frac{C_{3}}{B}r^{-2(1-\lambda)}.\label{r_dot_JMN2}\ee
Then we have
\be r\approx\left[(2-\lambda)\sqrt{\frac{C_{3}}{B}}\tau\right]^{\frac{1}{2-\lambda}}.\label{r_JMN2}\ee
Equation~(\ref{r_dot_JMN2}) implies that near the center $|\dot{r}|\gg1$, so Eq.~(\ref{connection_t_r}) can be reduced to
\be \dot{t}\approx\sqrt{\frac{B}{A}}\dot{r}
\approx\frac{\sqrt{C_{3}}}{\alpha}r^{-2(1-\lambda)}
\approx\frac{\sqrt{C_{3}}}{\alpha}\left[(2-\lambda)\sqrt{\frac{C_{3}}{B}}\tau\right]^{-\frac{2(1-\lambda)}{2-\lambda}}. \label{t_dot_JMN2}
\ee
Setting $t|_{\tau=0}=0$, the solution to the above equation can be written as
\be t\approx\frac{(2-\lambda)\sqrt{C_{3}}}{\alpha\lambda}\left[(2-\lambda)\sqrt{\frac{C_{3}}{B}}\right]^{-\frac{2(1-\lambda)}{2-\lambda}}\tau^{\frac{\lambda}{2-\lambda}}.
\label{t_JMN2}\ee

Substitution of Eq.~(\ref{r_JMN2}) into (\ref{solution_x}) generates
\be x\sim\tau^{-\frac{\lambda}{2-\lambda}}.\label{x_JMN2}\ee
Note that, for the JMN2 model, $B=\mbox{Const}$, and $\dot{r}^2\gg1$ near the center. Then Eq.~(\ref{equation_a}) can be simplified as
\be
\ddot{a}+\left(\frac{A'}{2A}\dot{r}\right)\dot{a}+\left[\frac{1}{2}\left(-\frac{A'^2}{A^2}+\frac{A''}{A}\right)\dot{r}^{2}\right]a\approx 0.
\label{equation_a_JMN2}
\ee
Note that near the center $A\approx\alpha^{2}r^{2(1-\lambda)}$ and also consider Eqs.~(\ref{r_dot_JMN2}) and (\ref{r_JMN2}), one can reduce Eq.~(\ref{equation_a_JMN2}) to
\be \ddot{a}+\left(\frac{1-\lambda}{2-\lambda}\tau^{-1}\right)\dot{a}+\left[\frac{-1+\lambda}{(2-\lambda)^2}\tau^{-2}\right]a\approx 0.
\label{equation_a_JMN2_v2}
\ee
Equation~(\ref{equation_a_JMN2_v2}) can be solved by the method of Frobenius~\cite{Bender_1978,Nolan:1999tw}. The indicial equation is
\be s(s-1)+s\frac{1-\lambda}{2-\lambda}+\frac{-1+\lambda}{(2-\lambda)^2}=0,\label{indicial_eq_JMN2}\ee
which roots are $s_{1,2}=(1\pm\sqrt{5-4\lambda})/(4-2\lambda)$. As discussed in Sec.~\ref{sec:conditions}, for the JMN2 model, $1/\sqrt{2}\le\lambda<1$. Then we have $s_{1}>0$ and $s_{2}<0$. Therefore, as the center is approached,
\be a\sim\tau^{\frac{1-\sqrt{5-4\lambda}}{4-2\lambda}}.\label{solution_a_JMN2}\ee
Substituting Eqs.~(\ref{r_JMN2}), (\ref{x_JMN2}) and (\ref{solution_a_JMN2}) into (\ref{volume}), one obtains
\be
\displaystyle{\lim_{\tau \to 0}}\parallel{V(\tau)}\parallel\sim\tau^{\frac{5-4\lambda-\sqrt{5-4\lambda}}{4-2\lambda}}
\sim r^{\frac{5-4\lambda-\sqrt{5-4\lambda}}{2}}\to 0,\label{volume_JMN2}\ee
which implies that the singularity is gravitationally strong.

Since $|\dot{r}|\gg1$ near the center, Eq.~(\ref{connection_t_r}) is reduced to $\dot{t}^2\approx B\dot{r}^2/A$. With this result and Eqs.~(\ref{Rtt})-(\ref{condition_strong_2}), we arrive at
\be \displaystyle{\lim_{\tau \to 0}}\tau^2 R_{ab}T^{a}T^{b}
=\displaystyle{\lim_{\tau \to 0}} \tau^2 (R_{tt}\dot{t}^2+R_{rr}\dot{r}^2)\approx\frac{A'}{Ar}\dot{r}^2\tau^2\approx\frac{2(1-\lambda)}{(2-\lambda)^2}>0.\label{condition_JMN2}\ee
So this also shows that the singularity is strong.

\begin{figure*}[t!]
  \epsfig{file=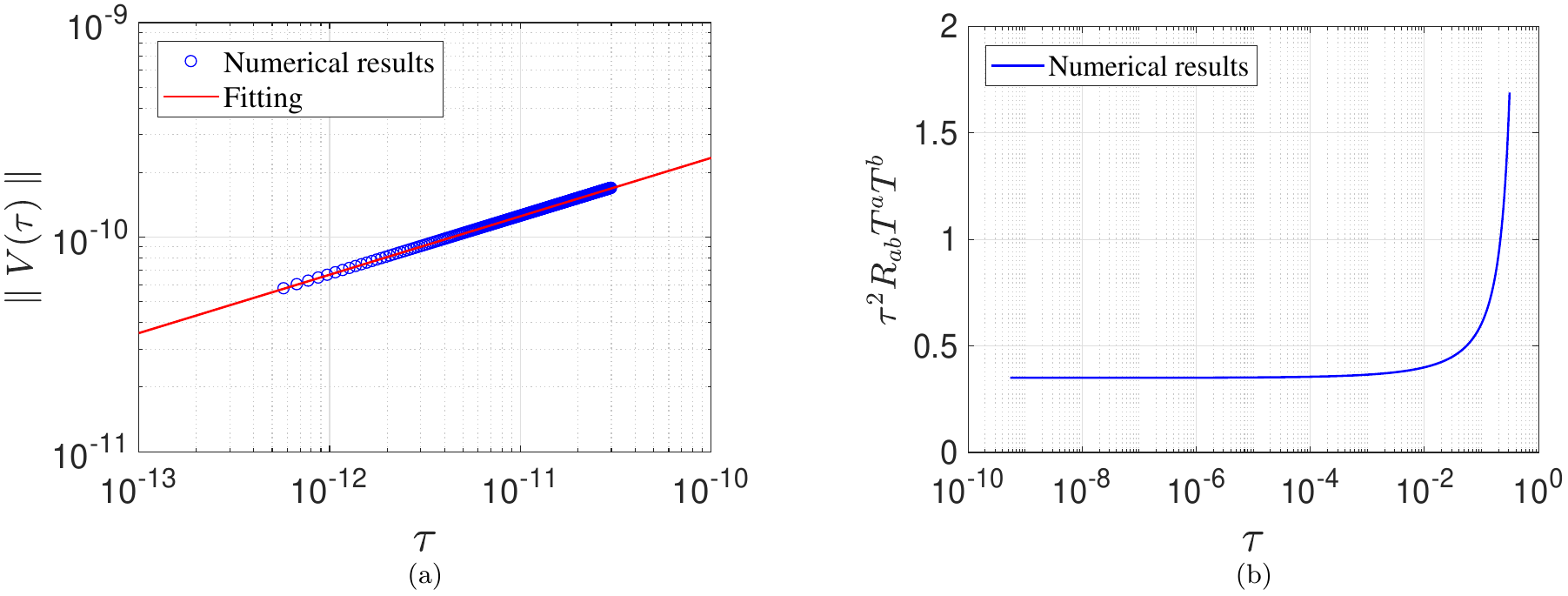,width=0.8\textwidth}
  \caption{Numerical results for the strength of the naked singularity in the JMN2 model with $\lambda=1/\sqrt{2}$. (a) The norm of the volume element $\parallel V(\tau)\parallel$ vs. $\tau$. $\displaystyle{\lim_{\tau\to 0}}\parallel V(\tau)\parallel\sim\tau^{(5-4\lambda-\sqrt{5-4\lambda})/(4-2\lambda)}$. See Eq.~(\ref{volume_JMN2}). For $\lambda=1/\sqrt{2}$, $(5-4\lambda-\sqrt{5-4\lambda})/(4-2\lambda)\approx0.26992$. We fit the numerical results of $\parallel V(\tau)\parallel$ vs. $\tau$ according to $\ln\parallel V(\tau)\parallel\approx a\ln\tau+b$, and obtain $a=0.27220\pm0.00001$ and $b=-15.9078\pm0.0003$. (b) $\tau^2 R_{ab}T^{a}T^{b}$ vs. $\tau$. $\displaystyle{\lim_{\tau \to 0}} \tau^2 R_{ab}T^{a}T^{b}\approx 2(1-\lambda)/(2-\lambda)^2$. See Eq.~(\ref{condition_JMN2}). For $\lambda=1/\sqrt{2}$, $2(1-\lambda)/(2-\lambda)^2\approx 0.35044$. The corresponding numerical result is $\displaystyle{\lim_{\tau \to 0}} \tau^2 R_{ab}T^{a}T^{b}\approx 0.35049$. The results plotted in (a) and (b) show that the naked singularity in the JMN2 model is gravitationally strong.}
  \label{fig:strength_JMN2}
\end{figure*}

The strength of the singularity are also investigated numerically. We numerically integrate Eqs.~(\ref{geodesic_r_1}), (\ref{geodesic_t}), (\ref{equation_x}) and (\ref{equation_a}) with proper initial conditions, and then compute the volume norm~(\ref{volume}) and $\displaystyle{\lim_{\tau \to 0}}\tau^2 R_{ab}T^{a}T^{b}$ along one geodesic toward the center. The initial conditions are the following with the {\lq}constraint{\rq} equation~(\ref{connection_t_r}) being satisfied: $r=0.02r_{b}$, $\dot{r}=-0.01$, $\dot{t}=3.7391$, $x=-1$, $\dot{x}=-0.01$, $a=0.5$ and $\dot{a}=-0.01$. The results by asymptotic analysis and numerical integration match well. Regarding the result (\ref{volume_JMN2}), in the case of $\lambda=1/\sqrt{2}$, $(5-4\lambda-\sqrt{5-4\lambda})/(4-2\lambda)\approx0.26992$. Fitting the numerical results of $\parallel{V(\tau)}\parallel$ vs. $\tau$ according to $\ln\parallel{V(\tau)}\parallel\approx a\ln\tau+b$, we obtain $a=0.27220\pm0.00001$ and $b=-15.9078\pm0.0003$. See Fig.~\ref{fig:strength_JMN2}(a). Regarding the result (\ref{condition_JMN2}), in the case of $\lambda=1/\sqrt{2}$, $2(1-\lambda)/(2-\lambda)^2\approx 0.35044$. The corresponding numerical result is $\displaystyle{\lim_{\tau \to 0}} \tau^2 R_{ab}T^{a}T^{b}\approx 0.35049$. See Fig.~\ref{fig:strength_JMN2}(b).

\subsection{Dehnen model}
For the Dehnen model, we first consider the case of $1\le\gamma<2$. In this case, near the center, there are
\be  A\approx e^{2\phi_0}\exp\left[\frac{M_{0}}{2-\gamma}\left(\frac{r}{r_{0}}\right)^{2-\gamma}\right],
\hphantom{dddd}B\approx1+{M_0}\left(\frac{r}{r_{0}}\right)^{2-\gamma}.
\label{A_B_Dehnen}\ee
Then the geodesic equation (\ref{geodesic_r_2}) is reduced to
\be \ddot{r}\approx-\frac{M_{0}}{2r_{0}^{2-\gamma}}r^{1-\gamma}[1+(3-\gamma)\dot{r}^2].\label{geodesic_r_Dehnen}\ee
Multiply both sides of Eq.~(\ref{geodesic_r_Dehnen}) by $2(3-\gamma)\dot{r}/[1+(3-\gamma)\dot{r}^2]$ and integrate, one obtains
\be \ln[1+(3-\gamma)\dot{r}^2]\approx-\frac{M_{0}(3-\gamma)}{r_{0}^{2-\gamma}(2-\gamma)}r^{2-\gamma}+C_{4}.\label{sln_r_Dehnen_1}\ee
So, as $r$ approaches zero, there are
\be \dot{r}\to \mbox{Const.,} \hphantom{dddd}r\sim\tau.\label{sln_r_Dehnen_2}\ee
Moreover, with Eqs.~(\ref{geodesic_t_2}) and (\ref{A_B_Dehnen}), near the center, there are also
\be \dot{t}\to \mbox{Const.,} \hphantom{dddd}t\sim\tau.\label{sln_t_dot_Dehnen}\ee
With Eqs.~(\ref{solution_x}) and (\ref{sln_r_Dehnen_2}), we arrive at
\be x\sim\frac{1}{\tau}.\label{x_Dehnen}\ee

According to Eq.~(\ref{A_B_Dehnen}), there are $(A'/A)^2\ll|A''/A|$ and $(B'/B)^2\ll|B''/B|$. Combining these results and Eq.~(\ref{sln_r_Dehnen_2}), one can rewrite Eq.~(\ref{equation_a}) as
\be \ddot{a}+ C_{5}\tau^{1-\gamma}\dot{a}\approx C_{6}\tau^{-\gamma}a.
\label{equation_a_Dehnen_2}
\ee
The numerical results show that, as $\tau$ approaches zero,
\be \dot{a}\sim\tau^{1-\gamma}, \hphantom{dddd}a(\tau)\to \mbox{Const}.\label{sln_a_Dehnen}\ee
Combination of Eqs.~(\ref{sln_r_Dehnen_2}), (\ref{x_Dehnen}), (\ref{sln_a_Dehnen}) and (\ref{volume}) yields
\be \displaystyle{\lim_{\tau \to 0}}\parallel V(\tau)\parallel\sim|a|>0.\label{volume_Dehnen}\ee
So the singularity for the Dehnen model with $1<\gamma<2$ is gravitationally weak.

For the $\gamma=1$ case, the numerical results also generate that near the center, $a\to\mbox{Constant}$ and $\parallel V(\tau)\parallel>0$. So the singularity is also weak. Details are skipped.

Near the center, $R_{tt}\sim A''/(2B)+A'/r\sim\tau^{-\gamma}$, and $R_{rr}\sim-A''/(2A)+B'/(Br)\sim\tau^{-\gamma}$. Combining these results and Eqs.~(\ref{sln_r_Dehnen_2}) and (\ref{sln_t_dot_Dehnen}), we arrive at
\be
\displaystyle{\lim_{\tau \to 0}}\tau^2 R_{ab}T^{a}T^{b}
=\displaystyle{\lim_{\tau \to 0}}\tau^2 (R_{tt}\dot{t}^2+R_{rr}\dot{r}^2)
\sim\tau^{2-\gamma}
{\sim}r^{2-\gamma}\to 0.
\label{condition_Dehnen}
\ee
So this also shows that the singularity for the Dehnen model with $1\le\gamma<2$ is weak.

\begin{figure*}[t!]
  \epsfig{file=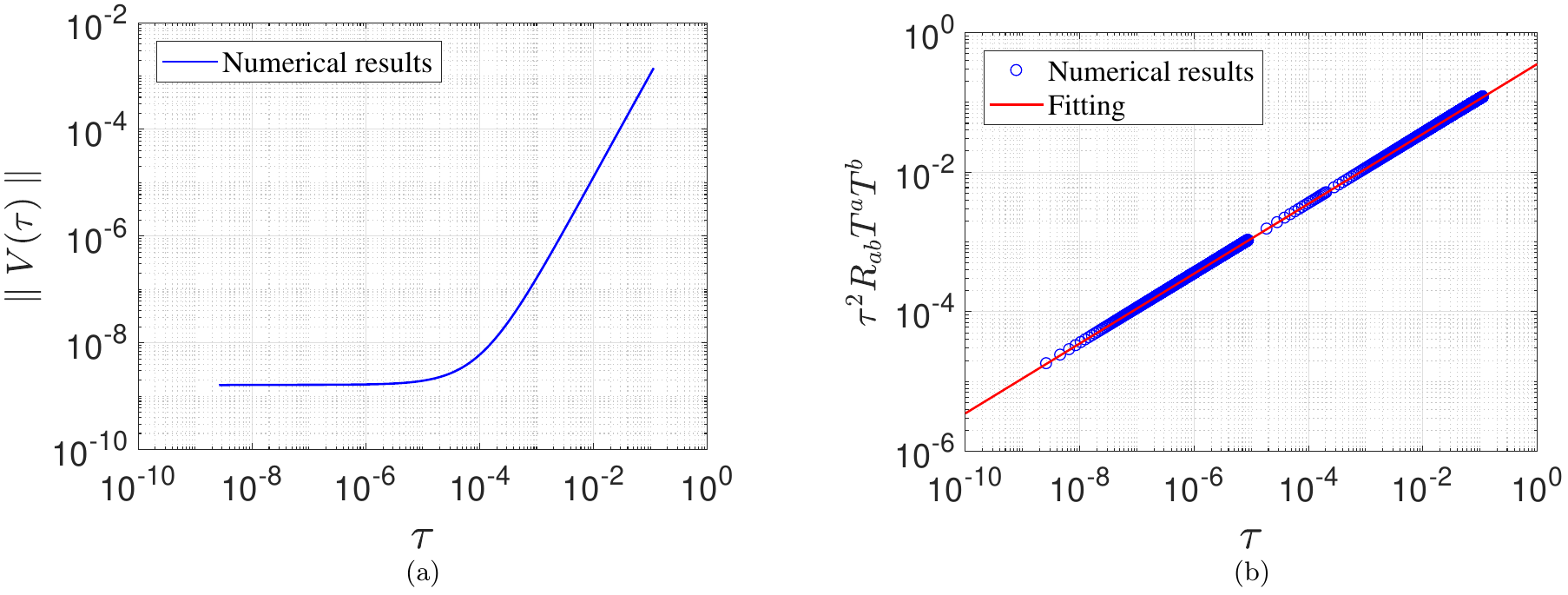,width=0.8\textwidth}
  \caption{Numerical results for the strength of the naked singularity in the Dehnen model with $\gamma=1.5$. (a) The norm of the volume element $\parallel V(\tau)\parallel$ vs. $\tau$. $\displaystyle{\lim_{\tau \to 0}}\parallel V(\tau)\parallel\sim |a|>0$. See Eq.~(\ref{volume_Dehnen}). (b) $\tau^2 R_{ab}T^{a}T^{b}$ vs. $\tau$.
  $\displaystyle{\lim_{\tau \to 0}}\tau^2 R_{ab}T^{a}T^{b}\sim\tau^{2-\gamma}$, see Eq.~(\ref{condition_Dehnen}). In the case of $\gamma=1.5$, $2-\gamma=0.5$. We fit the numerical results of $\tau^2 R_{ab}T^{a}T^{b}$ vs. $\tau$ according to $\ln(\tau^{2}R_{ab}T^{a}T^{b})\approx a\ln\tau+b$, obtaining $a=0.500310\pm0.000004$ and $b=-1.04202\pm0.00002$. The results plotted in (a) and (b) show that the naked singularity in the Dehnen model with $\gamma=1.5$ is gravitationally weak.}
  \label{fig:strength_Dehnen}
\end{figure*}

Similar to what has been done to the JMN2 model in the last subsection, the strength of the singularity for the Dehnen model with $\gamma=1.5$ is also studied numerically. The results match well with the analytic ones. The initial conditions in the numerical integrations are the following: $r=0.01r_{b}$, $\dot{r}=-0.9934$, $\dot{t}=2.3197$, $x=-1$, $\dot{x}=-0.01$, $a=0.1$ and $\dot{a}=-0.01$. The numerical results for $\parallel V(\tau)\parallel$ vs. $\tau$ for the Dehnen model with $\gamma=1.5$ plotted in Fig.~\ref{fig:strength_Dehnen}(a) confirm the analytical result by Eq.~(\ref{volume_Dehnen}). We fit the numerical results for $\tau^2 R_{ab}T^{a}T^{b}$ vs. $\tau$ in the case of $\gamma=1.5$ according to $\ln(\tau^{2}R_{ab}T^{a}T^{b})\approx a\ln\tau+b$, and obtain $a=0.500310\pm0.000004$, and $b=-1.04202\pm0.00002$. See Fig.~\ref{fig:strength_Dehnen}(b).

Here we make a few comments:

\begin{enumerate}[fullwidth,itemindent=0em,label=(\roman*)]
  \item The numerical results also show that Eq.~(\ref{equation_a_Dehnen_2}) can be further reduced to
  \be \ddot{a}\approx C_{6}\tau^{-\gamma}a.
  \label{equation_a_Dehnen_3}
  \ee
  It can be shown by another approach that $a$ is bounded. See Appendix~\ref{sec:evolution_a}. Currently, we do not know how to obtain this result by a similar approach starting from Eq.~(\ref{equation_a_Dehnen_2}), rather than from (\ref{equation_a_Dehnen_3}).
  \item We arrive at the solution~(\ref{sln_a_Dehnen}) only numerically, and currently we do not know how to come to it analytically.
  \item As discussed in Sec.~\ref{sec:luminosity}, near the center, the Dehnen model with $\gamma=2$ is reduced to the JMN2 model, and the corresponding singularity is strong. Near the center, the NFW model corresponds to the Dehnen model with $\gamma=1$, and the corresponding singularity is weak.
\end{enumerate}

\subsection{Kretschmann scalar}
The Kretschmann scalar $K$ is another effective quantity to depict the geometric features of the spacetime near singularities. For a Schwarzschild black hole, in the vicinity of the central singularity, $r\propto\tau^{2/3}$~\cite{Guo:2013dha}. Then we have
\be K_{\scriptsize{\mbox{Schw}}}{\propto}r^{-6}\propto\tau^{-4}.\label{K_Schw}\ee
For the metric~(\ref{metric}), the Kretschmann scalar can be expressed as
\be
\begin{split}
K=&\frac{8D^2(\phi_{,r})^2}{r^2} + \frac{4(1-D)^2}{r^4} + \frac{2(D_{,r})^2}{r^2} + 4D^2(\phi_{,r})^4 + 8D^2(\phi_{,r})^2\phi_{,rr} \\
& + 4DD_{,r}(\phi_{,r})^3 + 4D^2(\phi_{,rr})^2 + 4DD_{,r}\phi_{,rr}\phi_{,r} + (D_{,r})^2(\phi_{,r})^2.
\end{split}
\label{K_scalar}\ee
With Eqs.~[(\ref{A_JMN2}), (\ref{r_JMN2}), (\ref{K_scalar})] and  [(\ref{define_F}), (\ref{F_Dehnen}), (\ref{dphidr_Dehnen}), (\ref{sln_r_Dehnen_2}), (\ref{K_scalar})], one respectively obtains that for the JMN2 and Dehnen models, near the center,
\be K_{\scriptsize{\mbox{JMN2}}}\propto r^{-4}\propto\tau^{-4/(2-\lambda)},\label{K_JMN2}\ee
\be K_{\scriptsize{\mbox{Dehnen}}}\propto r^{-2\gamma}\propto\tau^{-2\gamma}.\label{K_Dehnen}\ee

Regarding the relations of $K$ vs. $\tau$ for Schwarzschild spacetime, JMN2 and Dehnen models, the index $2\gamma$ in Eq.~(\ref{K_Dehnen}) goes to 4 as $\gamma\to2$, while the index $4/(2-\lambda)$ in Eq.~(\ref{K_JMN2}) is less than 4. So it seems that $K_{(\scriptsize{\mbox{Dehnen},\gamma<2})}$ is stronger than $K_{\scriptsize{\mbox{JMN2}}}$ and $K_{(\scriptsize{\mbox{Dehnen},\gamma=2})}$. Moreover, with the expressions in terms of $\tau$ in Eqs.~(\ref{K_Schw}) and (\ref{K_Dehnen}), it seems that $K_{(\scriptsize{\mbox{Dehnen},\gamma\to 2})}$ diverges at the same speed as $K_{\scriptsize{\mbox{Schw}}}$ does. How to explain these? Such confusions may be related to a critical question: which quantity is more basic, $\tau$ or $r$?

Considering the relations of $K$ vs. $r$, $K_{\scriptsize{\mbox{JMN2}}}$ and $K_{(\scriptsize{\mbox{Dehnen},\gamma=2})}$ are stronger than $K_{(\scriptsize{\mbox{Dehnen},\gamma<2})}$, and weaker than $K_{\scriptsize{\mbox{Schw}}}$. This picture seems more natural. Note that the Kretschmann scalar describes one geometric feature (curvature) of spacetime. The definition of strength of singularities that we use in this paper originates from considerations of the tidal forces between nearby particles falling radially towards a singularity, and also only involves geometric objects~\cite{Tipler_1977}. Therefore, it seems more proper to express the Kretschmann scalar and strength of singularities in terms of the geometric quantity $r$ rather than the proper time $\tau$. With these arguments, we also think that $r$ may be more basic than $\tau$.

\subsection{Efficiencies for the accretion disks vs. strength of the naked singularities}
Equations~(\ref{condition_Dehnen}) and (\ref{K_Dehnen}) show that for the Dehnen model, as $\gamma$ increases, the singularity becomes stronger. On the other hand, from Fig.~\ref{fig:efficiency}, one obtains that as $\gamma$ increases, the efficiencies for the accretion disks become higher. So there is a strong connection between the strength of the singularities and the efficiencies for the accretion disks. In fact, integrating Eq.~(\ref{phi_r}) and taking into account Eqs.~(\ref{E_accretion}), (\ref{boundary_Dehnen}), (\ref{efficiency_energy_loss}), (\ref{condition_Dehnen}) and (\ref{K_Dehnen}), one arrives at the same conclusion: when the parameter $\gamma$ increases from $1$ to $2$, the radius of the sphere decreases, the sphere becomes more compact, the energy per unit mass $E$ at the center decreases, the efficiency increases, and the singularity becomes stronger. The results on the strength of the naked singularities and thermal properties of the accretion disks are summarized in Table~\ref{table:compare}.

\begin{table*}
\renewcommand{\arraystretch}{1.3}
\newcommand{\tabincell}[2]{\begin{tabular}{@{}#1@{}}#2\end{tabular}}
\begin{tabular}{c|c|c|c|c|c|c|c|c}
 \Xhline{0.8pt}
  {\tabincell{c}{Model}}
  &\tabincell{c}{Density near\\the center}
  &\tabincell{c}{$r$ vs. $\tau$}
  &\tabincell{c}{$\displaystyle{\lim_{\tau \to 0}}\parallel{V(\tau)}\parallel$}&\tabincell{c}{${\displaystyle{\lim_{\tau \to 0}}\tau^2} R_{ab}T^{a}T^{b}$}
  &\tabincell{c}{Kretschmann\\ scalar}
  &\tabincell{c}{Strength of\\ singularity}
  &\tabincell{c}{Efficiency \\of the disk}
  &\tabincell{c}{Slope of \\spectrum \\at high $\nu$}\\
  \Xhline{0.8pt}
  \tabincell{c}{Schw.\\BH}
  &\tabincell{c}{Vacuum}
  &\tabincell{c}{$r\propto\tau^{2/3}$}
  &\tabincell{c}{$\sim\tau^{1/3}\propto r^{1/2}\to 0$}
  &\tabincell{c}{Not applicable}
  &\tabincell{c}{$K\propto r^{-6}$\\ $\propto\tau^{-4}$}
  &\tabincell{c}{Strong}
  &\tabincell{c}{0.057}
  &\tabincell{c}{$-\frac{h\nu}{kT_{(\infty,\mbox{\scriptsize min})}}$}\\
  \hline
  \tabincell{c}{JMN2\\ and $\gamma=2$\\ of Dehnen}
  &\tabincell{c}{$\epsilon\propto r^{-2}$\\ $\propto\tau^{-2/(2-\lambda)}$}
  &\tabincell{c}{$r\propto\tau^{1/(2-\lambda)}$}
  &\tabincell{c}{$\sim\tau^{\frac{5-4\lambda-\sqrt{5-4\lambda}}{4-2\lambda}}$\\$\propto r^{\frac{5-4\lambda-\sqrt{5-4\lambda}}{2}}\to 0$}
  &\tabincell{c}{$\approx\frac{2(1-\lambda)}{(2-\lambda)^2}>0$}
  &\tabincell{c}{$K\propto r^{-4}$\\ $\propto\tau^{-4/(2-\lambda)}$\\ $\propto\epsilon^{2}$}
  &\tabincell{c}{Strong}
  &\tabincell{c}{1}
  &\tabincell{c}{$\frac{4(\lambda-1)}{3\lambda-1}$}\\
  \hline
  \tabincell{c}{{\tabincell{c}{$1\le\gamma<2$\\ of Dehnen}}}
  &\tabincell{c}{$\epsilon\propto r^{-\gamma}$\\ $\propto\tau^{-\gamma}$}
  &\tabincell{c}{$r\propto\tau$}
  &\tabincell{l}{$\sim|a|>0$}
  &\tabincell{c}{$\sim\tau^{2-\gamma}$\\ $\propto r^{2-\gamma}\to 0$}
  &\tabincell{c}{$K\propto r^{-2\gamma}$\\ $\propto\tau^{-2\gamma}$\\ $\propto\epsilon^{2}$}
  &\tabincell{c}{Weak}
  &\tabincell{c}{$<1$}
  &\tabincell{c}{$4\left(1-\frac{2}{\gamma}\right)$}\\
  \hline
  \tabincell{c}{NFW}
  &\tabincell{c}{$\epsilon\propto r^{-1}$\\ $\propto\tau^{-1}$}
  &\tabincell{c}{$r\propto\tau$}
  &\tabincell{c}{$\sim|a|>0$}
  &\tabincell{c}{$\sim\tau\propto r\to 0$}
  &\tabincell{c}{$K\propto r^{-2}$\\ $\propto\tau^{-2}$\\ $\propto\epsilon^{2}$}
  &\tabincell{c}{Weak}
  &\tabincell{c}{$<1$}
  &\tabincell{c}{$-4$}\\
  \Xhline{0.8pt}
\end{tabular}
\caption{Strength of the naked singularities and thermal properties of the accretion disks around the singularities. For the JMN2 model and the Dehnen model with $\gamma=2$, along the geodesic, as the center is approached, $\displaystyle{\lim_{\tau \to 0}}\parallel{V(\tau)}\parallel\to 0$ and ${\displaystyle{\lim_{\tau \to 0}}\tau^2} R_{ab}T^{a}T^{b}>0$. So the corresponding singularities are gravitationally strong. Moreover, the efficiencies for the conversion of the mass energy of the accreting gas into radiation are equal to $1$. For the Dehnen model with $1\le\gamma<2$ and the NFW model, $\displaystyle{\lim_{\tau \to 0}}\parallel{V(\tau)}\parallel>0$ and ${\displaystyle{\lim_{\tau \to 0}}\tau^2} R_{ab}T^{a}T^{b}\to 0$. So the corresponding singularities are gravitationally weak. The efficiencies are less than $1$. So the stronger the singularity is, the higher is the efficiency for the accretion disk. The slopes for the luminosity distributions, $\mbox{slope}{\equiv}d[\log_{10}(\nu {\cal L}_{\nu,\infty}/\dot{m})]/d[\log_{10}(h\nu/kT_{*})]$, at high frequencies take different expressions for different models. In principle, this feature can be used to distinguish the density profile models.}
\label{table:compare}
\end{table*}

\section{Concluding remarks\label{sec:summary}}
The thermal properties of the accretion disks around naked singularities, in some well-known density profile models for galaxies, and the gravitational strengths of singularities were investigated here. The spectral luminosity distributions for the accretion disks were computed in the corresponding cases. We obtained the asymptotic expressions for the slopes of the luminosity distributions at high frequencies. The results are significantly different from those for a Schwarzschild black hole of the same mass. Such differences can be used in principle, to distinguish naked singularities from black holes.

We studied the efficiencies for the conversion of the mass energy of the accreting mass into radiation. It was found that the efficiencies for the JMN2 model and the Dehnen model with $\gamma=2$ are equal to $1$; while for the Dehnen model with $1\le\gamma<2$ and the NFW model, not all the mass energy of the accreting gas can be converted into radiation.

Regarding the strength of the naked singularities, the singularities for the JMN2 model and the Dehnen model with $\gamma=2$ are strong, and these are weak for the Dehnen model with $1\le\gamma<2$ and the NFW model. It turns out that more compact the spheres for the models are, then much stronger are the singularities, and we have much higher efficiencies for the same.

We are aware that we work here on toy accretion disk models, in which the possible collisions between the test particles and the matter inside the sphere have been neglected. It would be interesting and useful to work on further generalizations of the models considered here.

\section*{Acknowledgments}
JQG is thankful to Dipanjan Dey and Karim Mosani for helpful discussions, and to the International Center for Cosmology, Charusat University and the Morningside Center of Mathematics, Academy of Mathematics and System Science, Chinese Academy of Sciences for hospitality. This work is supported by Shandong Province Natural Science Foundation under grant No.ZR2019MA068.

\begin{appendix}
\section{Derivations of the evolution equations for the Jacobi fields\label{sec:appendix_Jacobi_eq}}
In this appendix, we derive the evolution equations for the Jacobi fields~(\ref{equation_x}) and (\ref{equation_a}). We start with the geodesic deviation equations,
\be \frac{D^{2}{\xi}^a}{d\tau^2}+R_{cbd}^{\hphantom{ddd}a}\xi^{b}T^{c}T^{d}=0,\label{geodesic_deviation}\ee
where $D/d\tau[=(dx^{a}/d\tau)\nabla_{a}]$ is the directional covariant derivative, $\xi^a$ a Jacobi field, $R_{cbd}^{\hphantom{ddd}a}$ the Riemann tensor, and $T^c(\equiv dx^{c}/d\tau)$ a tangent vector to a radial timelike geodesic. Consider the Jacobi field
\be \overrightarrow{\xi}_{(1)}=x\frac{\partial}{\partial\theta}.\ee
Then in the metric~(\ref{metric}), there are
\be \frac{D\overrightarrow{\xi}_{(1)}}{d\tau}=\left(\dot{x}+\frac{x}{r}\dot{r}\right)\frac{\partial}{\partial\theta},\hphantom{dddd}
\frac{D^2\overrightarrow{\xi}_{(1)}}{d\tau^2}=\left(\ddot{x}+\frac{2\dot{r}\dot{x}}{r}+\frac{x}{r}\ddot{r}\right)\frac{\partial}{\partial\theta}.
\label{xi_dot2_x}
\ee
On the other hand, the nonzero components of $R_{c{\theta}d}^{\hphantom{ddd}\theta}$ are $R_{t{\theta}t}^{\hphantom{ddd}\theta}=A'/(2rB)$ and $R_{r{\theta}r}^{\hphantom{ddd}\theta}=B'/(2rB)$. So with Eq.~(\ref{geodesic_r_2}), we have
\be
R_{cbd}^{\hphantom{ddd}\theta}\xi^{b}_{(1)}T^{c}T^{d}
=R_{t{\theta}t}^{\hphantom{ddd}\theta}\xi^{\theta}_{(1)}\dot{t}^2+R_{r{\theta}r}^{\hphantom{ddd}\theta}\xi^{\theta}_{(1)}\dot{r}^2
=-\frac{x}{r}\ddot{r}.
\label{Riemann_theta}
\ee
Substitution of Eqs.~(\ref{xi_dot2_x}) and (\ref{Riemann_theta}) into (\ref{geodesic_deviation}) yields
\be r\ddot{x}+2\dot{r}\dot{x}=0,\ee
which is Eq.~(\ref{equation_x}).

Consider a Jacobi field in the radial 2-space,
\be \overrightarrow{\xi}_{(3)}=aB\dot{r}\frac{\partial}{\partial t}+aA\dot{t}\frac{\partial}{\partial r},\ee
which is orthogonal to the tangent vector to one radial timelike geodesic,
\be \overrightarrow{T}=\frac{d}{d\tau}=\dot{t}\frac{\partial}{\partial t}+\dot{r}\frac{\partial}{\partial r}.\ee
For the Jacobi field $\overrightarrow{\xi}_{(3)}$, with Eqs.~(\ref{geodesic_r_1}) and (\ref{geodesic_t_2}), there are
\beq
\frac{D\overrightarrow{\xi}_{(3)}}{d\tau}&=&\left[(aB\dot{r})^{\cdot}+(aB\dot{r})\frac{A'}{2A}\dot{r}+(aA\dot{t})\frac{A'}{2A}\dot{t}\right]\frac{\partial}{\partial t}
+\left[(aA\dot{t})^{\cdot}+(aA\dot{t})\frac{B'}{2B}\dot{r}+(aB\dot{r})\frac{A'}{2B}\dot{t}\right]\frac{\partial}{\partial r}\nonumber\\
&=&\eta\frac{\partial}{\partial t}+\sigma\frac{\partial}{\partial r},
\label{xi_dot_a}
\eeq
\be
\frac{D^2\overrightarrow{\xi}_{(3)}}{d\tau^2}=\left(\dot{\eta}+\eta\frac{A'}{2A}\dot{r}+\sigma\frac{A'}{2A}\dot{t}\right)\frac{\partial}{\partial t}
+\left(\dot{\sigma}+\sigma\frac{B'}{2B}\dot{r}+\eta\frac{A'}{2B}\dot{t}\right)\frac{\partial}{\partial r},
\label{xi_dot2_a}\ee
where
\be \eta=\dot{a}B\dot{r}+\frac{1}{2}a\dot{r}^2\left(\frac{A'B}{A}+B'\right), \hphantom{dddd} \sigma=\dot{a}A\dot{t}+\frac{1}{2}a\dot{r}\dot{t}\left(A'+\frac{AB'}{B}\right).
\label{define_eta_sigma}\ee

Using Eq.~(\ref{connection_t_r}) and the nonzero components of $R_{cbd}^{\hphantom{ddd}t}$ and $R_{cbd}^{\hphantom{ddd}r}$,
\be R_{rtr}^{\hphantom{ddd}t}=-R_{trr}^{\hphantom{ddd}t}=-\frac{1}{4}\left(\frac{2A''}{A}-\frac{A'^2}{A^2}-\frac{A'B'}{AB}\right),\hphantom{dddd} R_{trt}^{\hphantom{ddd}r}=-R_{rtt}^{\hphantom{ddd}r}=\frac{1}{4}\left(\frac{2A''}{B}-\frac{A'^2}{AB}-\frac{A'B'}{B^2}\right),\nonumber\ee
one obtains the $t-$ and $r$-components of $R_{cbd}^{\hphantom{ddd}a}\xi^{b}_{(3)}T^{c}T^{d}$,
\be
R_{cbd}^{\hphantom{ddd}t}\xi^{b}_{(3)}T^{c}T^{d}=\frac{1}{4}\left(\frac{2A''}{A}-\frac{A'^2}{A^2}-\frac{A'B'}{AB}\right)a\dot{r},\hphantom{dddd}
R_{cbd}^{\hphantom{ddd}r}\xi^{b}_{(3)}T^{c}T^{d}=\frac{1}{4}\left(\frac{2A''}{B}-\frac{A'^2}{AB}-\frac{A'B'}{B^2}\right)a\dot{t}.
\label{Riemann_t_r}
\ee

Substitution of Eqs.~(\ref{xi_dot2_a})-(\ref{Riemann_t_r}) into (\ref{geodesic_deviation}) yields
\be
\ddot{\xi}_{(3)}^a+R_{cbd}^{\hphantom{ddd}a}\xi_{(3)}^{b}T^{c}T^{d}=H\left(B\dot{r}\frac{\partial}{\partial t}+A\dot{t}\frac{\partial}{\partial r}\right)=0,
\ee
where
\be
H=\ddot{a}+\frac{1}{2}\dot{a}\dot{r}\left(\frac{A'}{A}+\frac{B'}{B}\right)+\frac{1}{2}a\left[\dot{r}^2\left(-\frac{A'^2}{A^2}-\frac{B'^2}{B^2}+\frac{A''}{A}+\frac{B''}{B}\right)
+\frac{1}{B}\left(-\frac{A'^2}{A^2}-\frac{A'B'}{AB}+\frac{A''}{A}\right)\right].
\ee
Then we arrive at Eq.~(\ref{equation_a}).

\section{Evolution of the quantity $a$ in $\mbox{Eq}$.~(\ref{equation_a_Dehnen_3})\label{sec:evolution_a}}
In this Appendix, we present another approach to solve Eq.~(\ref{equation_a_Dehnen_3}). Without loss of generality, we assume $a>0$ and let $a=e^{h}$. Then Eq.~(\ref{equation_a_Dehnen_3}) becomes
\be \ddot{h}+(\dot{h})^2\approx {C_{6}}\tau^{-\gamma}.\ee
We integrate the above equation with $\tau_0$ being fixed,
\be \int_{\tau}^{\tau_0}\ddot{h}d\tau + \int_{\tau}^{\tau_0}(\dot{h})^{2}d\tau\approx C_{6}\int_{\tau}^{\tau_0}\tau^{-\gamma}d\tau,\label{int_1}\ee
and obtain
\be \dot{h}|_{\tau_0} - \dot{h}|_{\tau}\le \frac{C_{6}}{1-\gamma}(\tau_{0}^{1-\gamma}-\tau^{1-\gamma}).\ee
Integrating the above equation and noting that $\tau_0$ is fixed, there is
\be \dot{h}|_{\tau_0}(\tau_0-\tau)-h|_{\tau_0} + h|_{\tau}\le \frac{C_{6}}{1-\gamma}\tau_{0}^{1-\gamma}(\tau_0-\tau)-\frac{C_{6}}{(1-\gamma)(2-\gamma)}\tau^{2-\gamma}|^{\tau_0}_\tau.\ee
Consequently, for $1<\gamma<2$, $h|_{\tau\to 0}$ and $a|_{\tau\to 0}$ are bounded.
\end{appendix}


\end{document}